\documentclass[reprint, prb]{revtex4-1}
\usepackage{amsmath}
\usepackage{amssymb}
\usepackage{graphicx}
\usepackage{enumitem}
\usepackage{bm}
\usepackage[caption=false]{subfig}
\usepackage{todonotes}
\usepackage{comment}
\graphicspath{{./}{./images/}}
\definecolor{darkblue}{rgb}{0.1,0.2,0.6} \definecolor{darkred}{rgb}{0.8,0.1,0.2}
\usepackage[colorlinks,citecolor=darkblue,linkcolor=darkred,urlcolor=darkblue]{hyperref}
\usepackage[all]{hypcap}
\renewcommand{\vec}[1]{\boldsymbol{\mathbf{#1}}}

\newcommand{\cf}{\textit{cf.} }

\newcommand{\etc}{\textit{etc.} }

\begin{document}

\title{Bipartite Fidelity and Loschmidt Echo of the Bosonic Conformal Interface}
 
\author{Tianci Zhou}
\email{tzhou13@illinois.edu}
\affiliation{University of Illinois, Department of Physics, 1110 W. Green St. Urbana, IL 61801 USA}

\author{Mao Lin}
\email{maolin2@illinois.edu}
\affiliation{University of Illinois, Department of Physics, 1110 W. Green St. Urbana, IL 61801 USA}

\date{\today}

\begin{abstract}

We study the quantum quench problem for a class of bosonic conformal interfaces by computing the Loschmidt echo and the bipartite fidelity. The quench can be viewed as a sudden change of boundary conditions parameterized by $\theta$ when connecting two one-dimensional critical systems. They are classified by $S(\theta)$ matrices associated with the current scattering processes on the interface. The resulting Loschmidt echo of the quench has long time algebraic decay $t^{-\alpha}$, whose exponent also appears in the finite size bipartite fidelity as $L^{-\frac{\alpha}{2}}$. We perform analytic and numerical calculations of the exponent $\alpha$, and find that it has a quadratic dependence on the change of $\theta$ if the prior and post quench boundary conditions are of the same type of $S$, while remaining $\frac{1}{4}$ otherwise. Possible physical realizations of these interfaces include, for instance, connecting different quantum wires (Luttinger liquids), quench of the topological phase edge states, \etc and the exponent can be detected in a X-ray edge singularity type experiment.



\end{abstract}

\maketitle

\section{Introduction}

In (one-dimensional) quantum critical systems, the presence of the physical boundary and isolated impurity weakly break the conformal symmetry. Simply put, the interface scatters the otherwise independent modes and therefore demonstrates novel boundary critical phenomena\cite{cardy_boundary_2004}. Operators close to the boundary are interpreted as boundary condition changing (bcc) operators\cite{oshikawa_boundary_1997,affleck_boundary_1997} in the boundary conformal field theory (CFT). Their correlation functions can exhibit different critical exponents from their bulk counterparts\cite{cardy_conformal_1984}. One example is the ``Anderson orthogonality catastrophe", where the core hole creates a potential that acts as an impurity to the conduction band. The X-ray absorption rate will then have a power law singularity of a boundary exponent\cite{affleck_boundary_1997} at the resonance frequency. There are numerous impurity problems of this kind that have been studied in the last few decades, such as the magnetic impurity in the spin chain\cite{eggert_magnetic_1992}, boundary and impurity effects in Luttinger liquid\cite{fabrizio_interacting_1995}, entanglement of the defects\cite{peschel_entanglement_2005, igloi_entanglement_2009,calabrese_entanglement_2012} \etc

\begin{figure}[h]
\includegraphics[width=1\columnwidth]{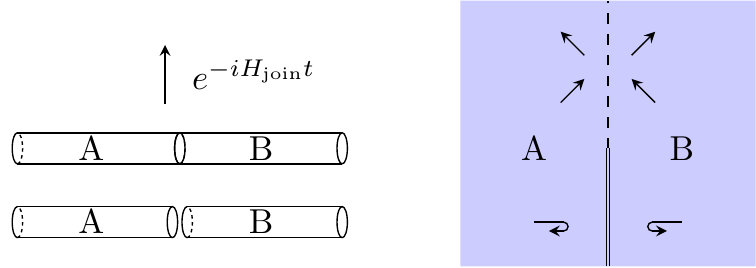}
\caption{Cut-and-join quench protocol. Left panel: Prepare the ground states of the two separated chains and join them at $t = 0$, then time evolve with the whole chain Hamiltonian. Right panel: Spacetime diagram of the cut and join protocol. The solid line represents the boundaries of the two disconnected chains. It is totally reflective for the incident particles on both sides. The dashed line is the world line of the junction, which we will call interface. It could either be totally transparent or partially permeable, depending the types of theories of A and B.}
\label{fig:cut-and-join}
\end{figure}

Recently, more attention has been paid to the non-equilibrium dynamics of quantum impurity\cite{hegde_quench_2015,francica_local_2016,lupo_transient_2016,lee_spatiotemporal_2016,chung_memory_2016,sacramento_edge_2016,vasseur_expansion_2015,mazza_overlap_2016}. The ``cut-and-join" quench protocol is a popular framework for investigating the spreading of the influence from the localized impurity (or boundary) across the system. As shown in the left panel of Fig.~\ref{fig:cut-and-join}, the system consists of two critical chains A and B, which were prepared in the ground states. They will be joined at $t = 0$ and evolve. Various quantities can be used to detect the information in the quench process. For instance, Ref.~\onlinecite{calabrese_entanglement_2007, calabrese_quantum_2016} find a logarithmic increase of entanglement entropy in subsystem A, when both A and B are identical critical systems. The authors ascribe such increase to the proliferation and propagation of the quasi-particle excitations emitted at the joint. Ref.~\onlinecite{vasseur_universal_2014} takes A to be a normal lead and B to be a topological superconductor in the topological phase. In this model, the Majorana zero mode acts as a bcc operator and its conformal dimension appears in the exponent of the power law decay of the Loschmidt echo.

In the path integral language, the ``cut-and-join" protocol corresponds to a spacetime diagram as shown in the right panel of Fig.~\ref{fig:cut-and-join}. The separating ground states prepared before $t = 0$ are joined to form a new type of interface between them. Before the quench, the slit represents boundaries that are completely reflective to the injecting particles. During the quench, the joining turns on the transmission from one side to the other. In the entanglement entropy and Loschmidt echo examples cited above\cite{calabrese_entanglement_2007, calabrese_quantum_2016, vasseur_universal_2014}, the two sides of the CFTs are the same (chiral fermion CFT in the case of Ref.~\onlinecite{vasseur_universal_2014}) and the boundary becomes totally transparent after the joining. 

In this paper, we generalize these ideas to an interface that interpolates between the totally reflective and complete transparent ones. This kind of interface can have many realizations. As discussed in Ref.~\onlinecite{bachas_permeable_2002}, one can connect two different bosonic CFTs in the ``cut-and-join" protocol, and the interface is a domain wall between two free compact boson theories with different compactification radii. Such permeable interface can also be implemented by non-compact free boson/fermion on a lattice with a fine-tuned bond interaction between the boundary sites (see \onlinecite{peschel_exact_2012,sakai_entanglement_2008} for their entanglement property studies). In these models, there is a parameter $\lambda$ that is directly related to the transmission coefficient. In the case of the compact boson, $\lambda$ is controlled by the ratio of the compactification radii, while for the free lattice boson it is controlled by the ratio of masses. We expect it to be tunable in a realistic experimental setting.

We compute the Loschmidt echo to extract information in the dynamics of the quench process of these models. The Loschmidt echo is the (square of the) overlap of the wavefunctions before the quench and the wavefunction evolved for some time $t$. It decays with a power law $t^{- \alpha}$ for the lack of length scale in the $t \rightarrow \infty$ limit. The decay exponent $\alpha$ has been calculated for various geometries and combinations of normal boundary conditions of the same CFTs in \onlinecite{stephan_logarithmic_2013,stephan_local_2011}. We extend the analysis to the aforementioned parametric interface of (possibly) different CFTs. We will see that there are two categories of the scattering matrices $S(\theta)$ of the interfaces, whose scattering angle parameter $\theta$ is determined by the transmission coefficients. Our analytic and numerical results show that $\alpha$ has a quadratic dependence on the change of $\theta$ if the prior and post quench boundary conditions are in the same type of $S$, while remaining $\frac{1}{4}$ otherwise. The finite size fidelity calculation further supports these results. 

The rest of the paper is organized as follows. In Sec.~\ref{sec:bosonic_conformal_interface}, we introduce the general formalism for the permeable bosonic conformal interface and its lattice realization. In Sec.~\ref{sec:analytic_numerics}, we analytically evaluate the free energy associated with the fidelity and Loschmidt echo, and present the numerical results for comparison. We discuss our results and related experimental works in Sec.~\ref{sec:disc}. Finally, we conclude in Sec.~\ref{sec:conclusion}. 

This paper includes several appendices for technical details. In App.~\ref{app:lambda_12}, we present the leading order analytical calculation of the free energy for the setups in Sec.~\ref{sec_sub:analy_eval}. In App.~\ref{app:gnd_dn_lambda}, we illustrate an alternative approach with one setup as an example. In App.~\ref{app:F_correction}, we point out two corrections to the free energy, which are complementary to the argument made in the main text. Up to this point, we work exclusively with the oscillator modes of the free bosons. In App.~\ref{app:compact_diff_boson}, it is shown that the winding modes of the compactified bosons will \emph{not} contribute to the free energy at the leading order. Therefore, the results remain valid in the physical situation of connecting two compactified bosons of different radii. In App.~\ref{app:pf_of_id}, we prove one identity that will be used repeatedly in the analytical evaluation. We derive the scale invariant interface for the free bosonic lattice in App.~\ref{app:interface_free_boson}. The details of the numerical simulation are presented in App.~\ref{app:comp_fid_echo}.



\section{Bosonic Conformal Interface}
\label{sec:bosonic_conformal_interface}

\subsection{General Formulation}
\label{sec_sub:general_formulation}

The general constraint on an interface is the continuity of the momentum flow across it. If we fold one side of the system on top of the other, then the resulting interface located on the boundary of the tensor theory (the crease of the folding) becomes impenetrable and the momentum flow should vanish there. This interface is naturally a conformal invariant boundary state\cite{cardy_boundary_2004,cardy_conformal_1984}. The interfaces in this paper are boundary states living in the $c = 2$ boundary CFT.

Although the general classification of the boundary states is still an open question\cite{affleck_quantum_2001}, there are many successful attempts to construct a subset of those bosonic boundary states. For example, one may use the current operator rather than the Virasoro generator to solve the zero-momentum flow condition. This idea dates back to the discovery of the Ishibashi state\cite{ishibashi_boundary_1989} and has been applied to the multi-component boson with a general compactification lattice\cite{affleck_quantum_2001,oshikawa_boundary_2010,quella_reflection_2007}. Additionally, the fusion algebra has also been used to generate new boundary states from the known ones, as shown in Ref.~\onlinecite{affleck_quantum_2001,bachas_fusion_2008}. 

We here follow the presentation in Ref.~\onlinecite{bachas_permeable_2002}, which imposes the conformal invariant boundary condition on the classical scalar fields and then quantize it to obtain the boundary state. The interface obtained is the same as the one by using the current algebra\cite{affleck_quantum_2001,oshikawa_boundary_2010,quella_reflection_2007}, but this viewpoint gives a more intuitive scattering picture and has more transparent relation to the discrete lattice model in Sec.~\ref{sec_sub:free_boson_lattice}. 

Assuming two free boson fields $\phi^1$ and $\phi^2$ living on the left and right half planes respectively, the interface located at $x = 0$ is characterized by the ``gluing condition"
\begin{equation}\begin{aligned}
\label{eq:def_M}
\begin{pmatrix}
\partial_t\phi^1\\
\partial_x\phi^1
\end{pmatrix}
=M\begin{pmatrix}
\partial_t\phi^2\\
\partial_x\phi^2
\end{pmatrix}.
\end{aligned}\end{equation}
The derivatives here should be understood in the appropriate left and right limits, for example $\partial_x \phi^1$ is evaluated at $ x = 0^-$. As argued before, the momentum components of the stress tensor is continuous across the interface. As a consequence $M$ is an element of the Lorentz group $O(1,1)$ and can be parameterized as
\begin{equation}\begin{aligned}
\label{eq:M1M2}
M_1(\theta)=\pm
\begin{bmatrix}
\lambda^{-1} & 0 \\
0 & \lambda
\end{bmatrix},\quad
M_2(\theta)=\pm
\begin{bmatrix}
0 & \lambda  \\
\lambda^{-1} & 0 
\end{bmatrix},
\end{aligned}\end{equation}
where $\lambda=\tan\theta$ for $\theta\in\left[-\frac{\pi}{2},\frac{\pi}{2}\right]$. 

Several special choices of $\theta$ need to be noted. 
\begin{enumerate}
\item $\theta=0,\pm \frac{\pi}{2}$. In this case, $\lambda$ (or $\lambda^{-1}$) appears to be singular and the field on either side of the interface cannot penetrate. The interface reduces to individual boundary conditions for the boson on the left and right half planes: They are a combination of the Dirichlet and Neumann boundary conditions. For example, $\lambda = 0$ for $M_1$ implies $\partial_x\phi^1 = \partial_t\phi^2 =0$, which means that the Dirichlet boundary condition is imposed on the right and the Neumann boundary condition on the left. Hereafter we shall denote this combination as `DN'. Similarly $M_1(\pm\frac{\pi}{2}),M_2(0),M_2(\pm \frac{\pi}{2})$ correspond to `ND', `DD', `NN' respectively. 
\item $\theta = \pm \frac{\pi}{4}$. In this case, $M_1(\theta)$ characterizes a perfectly transmitting interface. For example, there is effectively no interface in the case of $M_1( \frac{\pi}{4})$. We will denote it as ``P'' as it corresponds to the traditional periodic boundary condition. For the other three cases, despite picking up a phase, the two counter propagating modes are still fully transmitted across the interface. 
\end{enumerate}

The physical significance of $\theta$ will be clear in the scattering process described below. We rewrite Eq.~\eqref{eq:def_M} in the coordinates $t\pm x$ and use $\partial_{\pm} = \partial_{t} \pm \partial_x $ to extract the left and right going modes. For example, $\partial_{-} \phi^2$ will be a function of $t - x$ and hence represents a right going mode on the right half plane. This mode is one of the scattering modes that leave the interface. On the other hand, $\partial_{-} \phi^1$ and $\partial_{+} \phi^2$ are modes that approach the interface from their respective domains. We can therefore establish the scattering relation 
\begin{equation}\begin{aligned}
\label{eq:def_S}
\begin{pmatrix}
\partial_+\phi^1\\
\partial_-\phi^2
\end{pmatrix}
=S
\begin{pmatrix}
\partial_-\phi^1\\
\partial_+\phi^2
\end{pmatrix},
\end{aligned}\end{equation}
and solve the $S$ matrix for the two cases of $M_1$ and $M_2$
\begin{equation}\begin{aligned}
\label{eq:S1_S2}
S_1(\theta)=\begin{bmatrix}
\cos 2\theta & \sin 2\theta \\
\sin 2\theta & -\cos 2\theta
\end{bmatrix},\
S_2(\theta)=\begin{bmatrix}
-\cos 2\theta & \sin 2\theta \\
-\sin 2\theta & -\cos 2\theta
\end{bmatrix}.
\end{aligned}\end{equation}
For generic values of $\theta$, the interface is partially-transmitting, whose transmission coefficient is $\sin^2 2 \theta$.

We note that the $S$-matrices are independent of the wavelength, which agrees with the fact that the interface is scale invariant. 
\begin{figure}[h]
\centering
\includegraphics[width=0.45\textwidth]{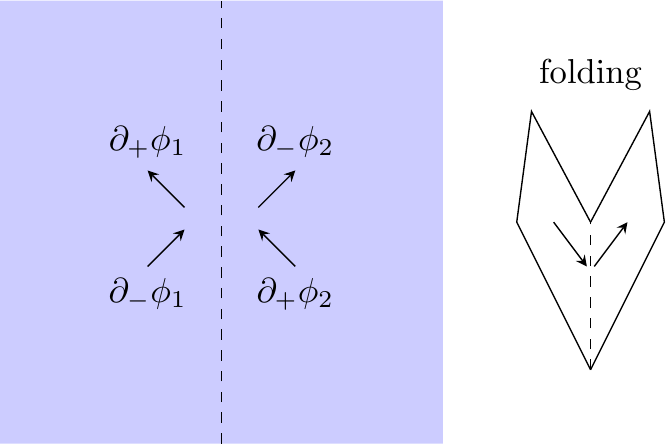}
\caption{Folding picture for the penetrable interface. Left panel: World line of the penetrable interface. $\partial_\pm\phi^{1,2}$ denote the left and right going modes in their respective domains. Right panel: Folding operation that sends $\phi^2(x)$ to $\phi^2(-x)$. The dashline represents the \emph{im}penetrable boundary for the resulting tensor theory. The arrow represents the incoming and outgoing particles scattered by the interface.}
\label{fig:folding_pic}
\end{figure}

We now work in the folding picture as shown in Fig.~\ref{fig:folding_pic}. The boundary at $x=0$ becomes impenetrable for the folded system, and the resulting tensor theory admits a conformal invariant boundary state. The folding sends $\phi^2(x)$ to $\phi^2(-x)$ and hence the gluing condition becomes
\begin{equation}
\begin{aligned}
\partial_t(\sin\theta\phi^1-\cos\theta\phi^2)=0, \quad
\partial_x(\cos\theta\phi^1+\sin\theta\phi^2)=0, 
\end{aligned}
\end{equation}
for the case $M = M_1(\theta)$. 

If we quantize the boson theory on the interface line $x = 0$, these gluing conditions become an identity for the boson creation and annihilation operators. We shall interpret these identities to be valid only when acting on the boundary states. The mode expansion of free boson at $x = 0$\cite{di_francesco_conformal_1997} is
\begin{equation}
\label{eq:di_mode_expansion}
\begin{aligned}
\phi(z, \bar{z} ) &= \phi_0 - \frac{i}{4\pi g } \pi_0 \ln z \bar{z} \\
 \quad&+ \frac{i}{\sqrt{4\pi g} } \sum_{n \ne 0 } \frac{1}{n} \left(a_n z^{-n} + \bar{a}_{n} \bar{z}^{-n}   \right),
\end{aligned}
\end{equation}
where we take the following choice of the holomorphic and anti-holomorphic coordinates
\begin{equation}
z= e^{ \frac{2\pi i(x - t)}{T} }, \quad \bar{z} = e^{ \frac{2\pi i( x + t  )}{T} },
\end{equation}
with $T$ being the time period. We end up with a set of operator identities for each mode
\begin{equation}
\begin{aligned}
\label{eq:rotation_a_basis}
\sin  \theta a^1_n - \cos \theta a^2_n  &= + \left( \sin  \theta \bar{a}^1_{-n} - \cos \theta \bar{a}^2_{-n} \right),\\
\cos  \theta a^1_n + \sin \theta a^2_n  &= - \left( \cos  \theta \bar{a}^1_{-n} + \sin \theta \bar{a}^2_{-n} \right), 
\end{aligned}
\end{equation}
which is valid for the following boundary state
\begin{equation}
\label{eq:bd_state}
\begin{aligned}
| B \rangle 
& =  \exp\Big\{ -\sum_{n > 0 } \frac{1}{n}
\begin{pmatrix}
a_{-n}^1\\
a_{-n}^2\\                              
\end{pmatrix}
S_1( \theta )
\begin{pmatrix}
\bar{a}_{-n}^1  \bar{a}_{-n}^2
\end{pmatrix} \Big\} |0\rangle,
\end{aligned}
\end{equation}
where $S_1(\theta) $ is precisely the scattering matrix in Eq.~\eqref{eq:def_S}. The calculation for the case $M=M_2(\theta)$ is completely analogous and we just have a replacement of $S_1( \theta ) $ by $S_2( \theta )$ in Eq.~\eqref{eq:def_S}. 

The boundary state expression in Eq.~\eqref{eq:bd_state} will be used extensively in the fidelity and Loschmidt echo calculation in Sec.~\ref{sec:analytic_numerics}. 

So far the derivation is only for the non-compact bosons, where the interface is determined by the ``gluing conditions''. For the case of connecting compact bosons of different radii, we will need to generalize the relation in Eq.~\eqref{eq:rotation_a_basis} to the winding mode operator $a_0$. Because the winding modes live on a compactification lattice, not all $\theta$ can satisfy Eq.~\eqref{eq:rotation_a_basis} for $a_0$. App.~\ref{app:compact_diff_boson} reviews the derivation about how the winding modes constrain the choice of $\theta$. For example, the $S_1(\theta)$ interface should satisfy
\begin{equation}
\lambda = \tan \theta = \frac{n_2 R_1}{n_1 R_2}
\end{equation}
for coprime integer $n_1$ and $n_2$ and compactification radii $R_1$ and $R_2$ for the bosons on the two sides. This also suggests that connecting two different CFTs will generate an interface whose transmission coefficient are determined by the universal parameters of the CFTs on both sides. 

The winding modes however do not contribute to the fidelity and echo exponent to the leading order, as shown in App.~\ref{app:compact_diff_boson}. So the derivations with the non-compact boson boundary state in Eq.~\eqref{eq:bd_state} holds true for the compact bosons.



\subsection{A Free Boson Lattice Model}
\label{sec_sub:free_boson_lattice}

In this section, we consider a lattice model with bosonic interface at the center\cite{peschel_exact_2012,calabrese_entanglement_2012}, which reduces to the one considered in Sec.~\ref{sec_sub:general_formulation} in the continuum limit\cite{sakai_entanglement_2008}. Therefore, it serves as a numerical tool to check our analytic results in Sec.~\ref{sec:analytic_numerics}.

We consider two harmonic chains with bosonic field $\phi_i$ and conjugate momentum $\pi_i$ at the lattice site $i$. The left and right chains are connected between site $0$ and $1$ with the Hamiltonian
\begin{equation}
\begin{aligned}
\label{eq:lattice_H}
H &= \frac{1}{2} \sum_i \pi_i^2  +  \frac{1}{2} \sum_{i\ne 0 }  ( \phi_i - \phi_{i+1} )^2 \\
\quad & + \frac{1}{2} \begin{pmatrix}  \phi_0, \phi_1 \end{pmatrix}
\begin{bmatrix}
1 + \Sigma_{11}  & \Sigma_{12} \\
\Sigma_{21} &  1 + \Sigma_{22} \\
\end{bmatrix}
\begin{pmatrix}
\phi_0 \\
\phi_1 
\end{pmatrix},
\end{aligned}
\end{equation}
where the $2\times2$ matrix $\Sigma$ parameterizes the two-site interaction. We performed the standard scattering analysis in App.~\ref{app:interface_free_boson}. For modes with momentum $k$, the only possible scale invariant $S$-matrix is
\begin{equation}
S = - e^{ika} \Sigma,
\end{equation}
where $a$ is the lattice constant. In the continuum limit $a\rightarrow0$, the matrix $\Sigma$ can be parameterized as
\begin{equation}
\label{eq:def_Sigma}
\Sigma = -\lim_{a \rightarrow 0 } S = 
\begin{bmatrix}
\frac{\lambda^2- 1}{1 + \lambda^2} & \frac{-2\lambda }{1 + \lambda^2} \\
\frac{-2\lambda }{1 + \lambda^2} & \frac{1- \lambda^2}{1 + \lambda^2} \\
\end{bmatrix},
\end{equation}
where $\lambda\in\mathbb{R}$. 

The lattice model in Eq.~\eqref{eq:lattice_H} with the bond interaction defined in Eq.~\eqref{eq:def_Sigma} will be used to check our analytic results for both the Loschmidt echo and the fidelity.



\section{Bipartite Fidelity and Loschmidt Echo}
\label{sec:analytic_numerics}

\subsection{Definition}

In this section, we define the fidelity and Loschmidt echo and present their corresponding imaginary time path integral diagrams.  We will see that these path integrals are just the free energy of boson with conformal interfaces (or boundaries). 

\begin{figure}[h]
\includegraphics[width=1\columnwidth]{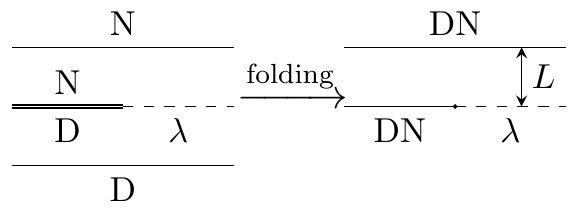}
\caption{Fidelity of connecting two CFTs. The horizontal axis is the imaginary time. Evolution along the two semi-infinite stripes produces the groundstates of the disconnected and connected chain Hamiltonians. The right diagram is the result of folding the lower part of the diagram up, so that all the boundaries are now boundary states. The solid dot represents boundary condition changing (bcc) operator. Here D(Dirichlet), N(Neumman) and $\lambda$(permeable interface parameterized by $\lambda$) are possible choices of boundary conditions.}
\label{fig:fidel}
\end{figure}

Fidelity is the square of the overlap of the groundstates of the two Hamiltonians, 
\begin{equation}
{\rm fidelity} \equiv |\langle \psi_1 |\psi_2  \rangle |^2 .
\end{equation}
For the systems we considered, $|\psi_1 \rangle$ is the groundstate of the two disconnected chains (of equal length $L$, hence ``bipartite") and $|\psi_2\rangle$ is that of the connected chains with the conformal interface. Both of them can be produced by an imaginary time evolution. Taking the horizontal axis as imaginary time direction, the fidelity can be diagrammatically represented in Fig.~\ref{fig:fidel}, where the slits represent the disconnected boundary conditions, such as Dirichlet(D) and Neumann(N), and the dashed line represents the conformal interface parameterized by $\lambda$. The logarithmic fidelity is then (twice) the free energy of this diagram
\begin{equation}
\mathcal{F}( {\rm fidelity} )  = - \ln \langle \psi_1 |\psi_2 \rangle^2 = - 2 \ln |Z|.
\end{equation}

\begin{figure}[h]
\centering
\includegraphics[width=1\columnwidth]{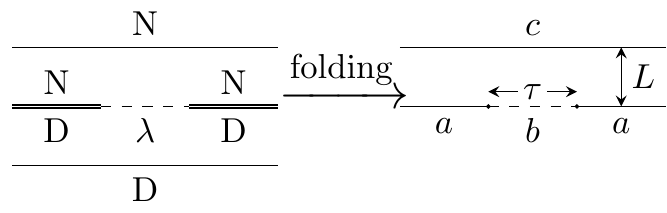}
\caption{Loschmidt echo of connecting two CFTs. Evolution along the two infinitely extended sides produces the groundstate of the disconnected chain Hamiltonians. They sandwich the evolution of the connected chains. In the folding picture on the right, $a,b,c$ represent the most general boundary conditions of the chains (for example, $a$ and $c$ are DN according to the left figure).}
\label{fig:echo}
\end{figure}

The Loschmidt echo is also (square of) the overlap of the two wavefunctions. One of them is the groundstate of the disconnected chains and the other is the groundstate evolved by the Hamiltonian of the connected chains
\begin{equation}
\mathcal{L}( t )  \equiv |\langle \psi_{\rm gnd}  | e^{-i H t } | \psi_{\rm gnd} \rangle|^2 .
\end{equation}
The imaginary time version $\mathcal{L}( \tau  ) = |\langle \psi_{\rm gnd}  | e^{-H \tau } | \psi_{\rm gnd} \rangle|^2$ has a path integral definition similar to Fig.~\ref{fig:cut-and-join}, but to be consistent with the fidelity diagram, we take the horizontal axis as imaginary time and present it in Fig.~\ref{fig:echo}. Viewing the diagram as a partition function subject to the switching of boundary conditions, the logarithmic Loschmidt echo is also the associated free energy. After obtaining the free energy in imaginary time, we can analytically continue back to real time to get the $t$ dependence. For simplicity and comparison with the fidelity result, we will take the length of both of the chains to be $L$ and set $L \gg t$, leaving $t$ the only length scale in the echo calculation (Fig.~\ref{fig:H-tau_fold}). The dependence on nonzero $\frac{t}{L}$ and the asymmetry of the lengths of the chains will not be discussed here (see the treatment in Ref.~\onlinecite{stephan_local_2011} for special values of $\lambda$).

If the interface is completely transparent, i.e. at the special point of $\lambda = 1$, the tip of the slit can be regarded as a corner singularity. According to Cardy and Peschel\cite{cardy_finite-size_1988}, the singularity will contribute a term that is logarithmic of the corner's characteristic size, which is $\ln L$ in the fidelity and $\ln \tau$ in the Loschmidt echo. One would expect the fidelity and echo to have power law decay with respect to these scales in the long wavelength limit. In fact, the computations have been done in Ref.~\onlinecite{stephan_logarithmic_2013,stephan_local_2011,vasseur_universal_2014,vasseur_crossover_2013,kennes_universal_2014} using either the Cardy-Peschel formula or the integral version of the Ward identity. If the slits boundary conditions are taken to be Dirichlet, we have the universal behavior for the leading term \cite{stephan_logarithmic_2013,stephan_local_2011}
\begin{equation}
\label{eq:full-pass}
\begin{aligned}
  \mathcal{F}({\rm fidelity}) &=  \frac{c}{8} \ln L, \\
 \mathcal{F}( {\rm echo} )  &= \frac{c}{4} \ln |\tau| \rightarrow \frac{c}{4} \ln |it + \epsilon|   \sim \frac{c}{4}\ln t ,
\end{aligned}
\end{equation}
where we have performed analytic continuation $\tau \rightarrow it + \epsilon$ with $\epsilon \rightarrow 0$ for the echo.

With the presence of the conformal interface, the tip of the slit is no longer a corner singularity\cite{cardy_finite-size_1988} . Its nature is clearer in the folding picture shown in Fig.~\ref{fig:fidel} and Fig.~\ref{fig:echo} where the lower half plane is flipped up on top of the upper half plane on both the fidelity and echo diagrams. From the boundary CFT point of view, the change of boundary conditions can be regarded as inserting a bcc operator. The diagrams for the fidelity and echo then become the one point or two point functions of the bcc operators respectively, and the free energy's leading logarithmic term extracts their scaling dimensions.



\subsection{Notation of Boundary Conditions}
\label{sec:notation}

We use chemical reaction style to represent the change of boundary conditions. Taking the example of the echo diagrams in Fig.~\ref{fig:echo}, there are three boundary conditions $a,b,c$ in the folding picture, which represents the status of the two ends of the chain before and after the quench. The choice of a uniform $c$ boundary condition on the far end of the chain is to isolate the effect coming from the bcc on the $ab$ interface. The process $a + c \rightarrow b + c$ represents the change of boundary condition from the combination $a$/$c$ prior to the quench to $b$/$c$ after the quench. Since each letter can take a general conformal interface defined by the $S$ matrix in Eq.~\eqref{eq:S1_S2}, we denote it as
\begin{equation}
\label{eq:Full_notation_rand()}
S_a( \theta_a ) + S_c( \theta_c) \rightarrow S_b( \theta_b )  + S_c( \theta_c ) .
\end{equation}
In most cases of the following, we will consider taking $a = c$ to remove the bcc operator from $a$ to $c$ at infinity. And we will use the shorthand notation
\begin{equation}
S_a( \theta_a ) \rightarrow S_b( \theta_b )
\end{equation}
to remind ourselves that we are isolating the boundary condition change only on the joint of the two chains. 

In the ``cut-and-join" protocol we considered, $a$ should be one of `DD', `DN', `ND', `NN', $b$ is taken to be $S_1( \theta )$ or $S_2( \theta )$. The physical situation of connecting two compact bosons (and our numerical simulation) corresponds to the choice of $S_1( \theta)$, and we reserve the notation $\lambda$ for this type of the boundary condition. For instance, the notation for the process presented in Fig.~\ref{fig:echo} is
\begin{equation}
\label{eq:DNDN}
 {\rm DN} \rightarrow \lambda.
\end{equation}
Another interesting case is to take $a$ or $c$ to be a completely transmitting interface, i.e. $S_2( \frac{\pi}{4} )$. This $S$-matrix corresponds to the traditional periodic boundary condition and we use symbol 'P' to denote it.



\subsection{Analytic Evaluation}
\label{sec_sub:analy_eval}

In this subsection, we relate the free energy to the amplitudes between the boundary states, and present the analytic results. 

We notice that there is only one apparent length scale in these diagrams -- the finite size $L$ for fidelity and imaginary time $\tau$ for the Loschmidt echo. These are the characteristic size of the corners at the tip of the slits. Regulators are necessary in keeping track of the scale dependence, otherwise a dilation transformation can rescale both $L$ and $\tau$ to $1$ and drop those scales. The introduction of the regulators is also physically sensible when considering the lattice realization of the systems. 

We thus add small semi-circles around the points where the bcc operators reside, and then apply a series of conformal mappings. 

\begin{figure}[h]
\centering
\includegraphics[width=\columnwidth]{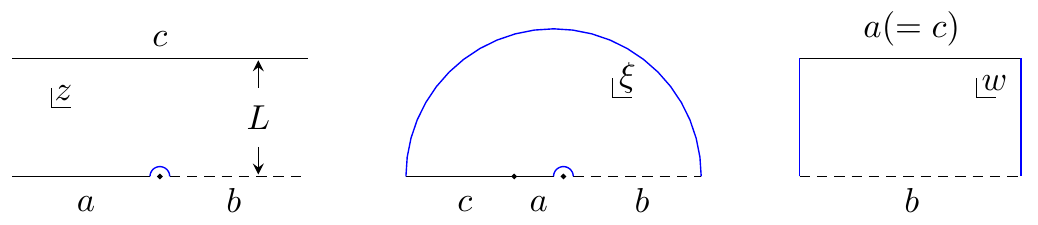}
\caption{Mapping from a strip to the upper half plane $\xi  = \exp( \frac{\pi z}{L} ) $. The two black dots represent possible locations of the boundary condition changing (bcc) operators. The dot inside the blue semi-circle has coordinate $\xi = 1$, which is the image of the point connecting $a$ and $b$ boundaries. The other dot $\xi = 0$ corresponds to the connection between $a$ and $c$ boundaries at $- \infty$. To evaluate the diagram, we add the outer blue semi-circle centered at $\xi = 1$ with radius $R_{\xi}$ to be the IR cut-off and map it to the cylinder with $w = \ln(\xi - 1)$}
\label{fig:fidel-map}
\end{figure}

For the fidelity case, the regulators as well as the conformal maps are depicted in Fig.~\ref{fig:fidel-map}. 
We add a small blue semi-circle to the folded strip in Fig.~\ref{fig:fidel} as the UV regulator and map it to the upper half plane using $\xi  = \exp( \frac{\pi z}{L} )$. Then both $\xi = 0$ and $1$ can host bcc operators. We assume $a = c$ such that the only bcc operator on the real axis is the one enclosed by the blue semi-circle around $\xi = 1$. In order to evaluate this diagram, we add another semi-circle centered around $\xi = 1$ with radius $R_{\xi}$ (this will introduce a correction as explained in App.~\ref{app:F_correction}), and map it to a cylinder of height $\pi$ on the right by $w = \ln ( \xi- 1)$. Finally the cylinder diagram can be viewed as an imaginary time path integral amplitude between the boundary states $b$ and $a$
\begin{equation}
\label{eq:partition_fun}
Z_{ab} = \langle a | e^{-\pi H } |b \rangle.
\end{equation}
The two end points of the $\epsilon$ radius semi-circle on the $z$ plane are mapped to
\begin{equation}
\exp( \pm \pi \frac{\epsilon}{ L}  ) \sim 1 \pm \pi \frac{\epsilon}{L} .
\end{equation}
The bigger blue semi-circle intersects the real axis at $1 \pm R_{\xi}$ and so the width of the cylinder is 
\begin{equation}
\label{eq:fidel_cyd_width}
\ln R_{\xi} - \ln \frac{\pi \epsilon}{L} = \ln L + \text{constant}.
\end{equation}

\begin{figure}[htb]
\centering
\includegraphics[width=\columnwidth]{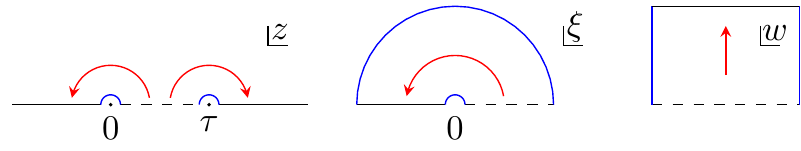}
\caption{The dashed (solid) lines are gluing (completely reflective) boundary conditions. Red arrows are the directions of Hamiltonian flow that propagates the dashed line boundary state to the solid line boundary state. Left: Diagram of the Loschmidt echo that reduces to a partition function with imaginary time in the horizontal direction. The blue semi-circles of radius $\epsilon$ are the UV regulators and they are identified as periodic boundaries in the direction perpendicular to the red arrow (equal time slice). Middle: Image of the map $\xi = \frac{z}{\tau - z}$. The two semi-circles have radii $({\tau}/{\epsilon})^{\pm1}$ respectively.  Right: Image of $w = \ln \xi$. It is a cylinder by identifying the blue lines and the standard radial quantization procedure can be applied. }
\label{fig:H-tau_fold}
\end{figure}

The Loschmidt echo can be evaluated in the same way. Again, we introduce two semi-circles (blue in Fig.~\ref{fig:H-tau_fold}) as regulators and then perform the conformal transformation shown in Fig.~\ref{fig:H-tau_fold}. From the $z$ plane to the $\xi$ plane, we use $\xi = \frac{z}{\tau - z}$ to map the two slits to half of an annulus, which is the same as the fidelity case. With one more conformal mapping $w = \ln \xi$, the diagram again becomes the cylinder partition function between the two boundary states. 

The height of the cylinder is still $\pi$. In the $\xi$ plane, the coordinates of the two end points of the small semi-circle are $\frac{\pm \epsilon}{ \tau \mp \epsilon} \sim \frac{\pm \epsilon}{ \tau }$, while those of the larger semi-circle are $\mp \frac{\tau \pm \epsilon}{\epsilon} \sim \frac{\mp \tau}{\epsilon}$. Hence the width of the cylinder is
\begin{equation}
\label{eq:echo_cyd_width}
\ln \frac{\tau}{\epsilon} - \ln \frac{\epsilon}{\tau } = 2 \ln \tau + \text{constant}.  
\end{equation}

One subtlety of the above description is that the two semi-circles in the center diagrams of Fig.~\ref{fig:fidel-map} and Fig.~\ref{fig:H-tau_fold} are not precisely concentric. This can be resolved by the following observation. There exists a conformal map $\zeta(\xi)$ that maps the non-concentric circles to the two standard concentric circles of radii $1$ and $R$ ($R>1$) on the $\zeta$ plane\cite{brown_complex_2009}. Then the logarithmic map $w = \ln \zeta$ produces a cylinder of width $\ln R$. In our case, since the height of the cylinder is always $\pi$, the width of the cylinder is a conformal invariant that only depends on the cross ratio of the half annulus. The four intersection points of two standard concentric circles on the $\zeta$ plane are $(\pm 1,0)$ and $(\pm R,0)$, whose cross ratio is
\begin{equation}
\eta = \frac{(1 + R)^2}{(1 - R)^2}. 
\end{equation}
Hence the width of the cylinder is $\ln \frac{\sqrt{ \eta } - 1}{\sqrt{ \eta} + 1}$. Since conformal transformation preserves the cross ratio, the result is the same if we use the cross ratio of the slightly non-concentric diagrams of Fig.~\ref{fig:fidel-map} and Fig.~\ref{fig:H-tau_fold}. The calculation in Eq.~\eqref{eq:fidel_cyd_width} and Eq.~\eqref{eq:echo_cyd_width} equivalently use the leading order approximation to $\eta$ in the respective geometries and thus get the leading order term in the width of the cylinders. The slight deviation to the precise concentric geometry will only bring in $\frac{\epsilon}{L}, \frac{\epsilon}{\tau}$ corrections to $\eta$ and the width parameter, which will not affect the fidelity and echo exponents. 

For the rest of this section, we should denote the width of the cylinder as $\beta$. After obtaining the partition function on it, we should set $\beta = 2 \ln L$ or $ 4 \ln \tau$ because the fidelity and Loschmidt echo are both square of the amplitudes.

The actual boundary conditions on the blue lines, which are the regulators in Fig.~\ref{fig:fidel-map} and Fig.~\ref{fig:H-tau_fold}, are not important in the leading order. Taking Fig.~\ref{fig:fidel-map} for example, rather than using Eq.~\eqref{eq:partition_fun}, we can alternatively view the right panel as the amplitude between the two blue boundary states $|1\rangle$ and $|2\rangle$
\begin{equation}
Z_{ab} =  \langle 1 | e^{ - \beta H_{ab}} | 2 \rangle  , 
\end{equation}
where $H_{ab}$ is the Hamiltonian with boundary condition $a$ and $b$. Since $\beta$ is taken to be large, we expect the imaginary time evolution (which is horizontal in this case) to project out only the groundstate $|0_{ab}\rangle $. Hence the free energy is
\begin{equation}
\begin{aligned}
F &=  - \ln Z_{ab}  \sim - \ln \langle 1 |0_{ab} \rangle \langle   0_{ab}  |e^{ - \beta H_{ab}}|0_{ab} \rangle \langle 0_{ab} | 2 \rangle \\
&= \beta  E_c - \ln \langle 1| 0_{ab} \rangle  - \ln \langle 0_{ab}  |2 \rangle,
\end{aligned}
\end{equation}
where $E_c$ is the groundstate/Casimir energy of $H_{ab}$. We see that different choices of the boundary conditions only change the term independent of $\beta$. Thus in the leading order we can choose any boundary conditions. The one we pick is the simplest one: the periodic boundary condition that identifies the two blue lines. 

With these simplifications, we now set up the partition function calculation of the general process $S_a( \theta_1 ) \rightarrow S_b( \theta_2)$. We define a set of bosonic operators related to the $a^i_n$s in Eq.~\eqref{eq:di_mode_expansion} through
\begin{equation}
\begin{aligned}
b^i_n &= \frac{a^i_n}{\sqrt{n}} \quad (b^i_n)^{\dagger} &= \frac{a^i_{-n}}{\sqrt{n}} \\
\bar{b}^i_n &= \frac{\bar{a}^i_n}{\sqrt{n}} \quad (\bar{b}^i_n)^{\dagger} &= \frac{\bar{a}^i_{-n}}{\sqrt{n}} \\
\end{aligned}
\end{equation}
for $n > 0 , i = 1, 2$, and group them compactly with the vector notation
\begin{equation}
\begin{aligned}
\vec{b}_i &= ( b^i_1, b^i_2, \cdots )^\top, \qquad \,\,\quad \vec{\bar{b}}_i = ( \bar{b}^i_1, \bar{b}^i_2, \cdots )^\top\\
\vec{b}^\dagger_i &= ( (b^i_1)^\dagger, (b^i_2)^\dagger, \cdots )^\top, \quad\vec{\bar{b}}^{\dagger}_i = ( (\bar{b}^i_1)^\dagger, (\bar{b}^i_2)^\dagger, \cdots )^\top.
\end{aligned}
\end{equation}
The boundary state in Eq.~\eqref{eq:bd_state} is then 
\begin{equation}
\exp\Big\{  (\vec{b}_1^{\dagger} \vec{b}_2^{\dagger} ) R_a( \theta )   
\begin{pmatrix}
  \vec{\bar{b}}_1^{\dagger}\\
  \vec{\bar{b}}_2^{\dagger}
\end{pmatrix}\Big\}  |0  \rangle ,
\end{equation}
where $R_a( \theta ) = -S_a \otimes \mathbb{I}$. Using even a lazier notation $\vec{b} = ( \vec{b}_1, \vec{b}_2 )^\top$, we have
\begin{equation}
\label{eq:bd_state_matrix}
|a \rangle = \exp\Big\{  \vec{b}^{\dagger} R_a( \theta )    \vec{\bar{b}}^{\dagger} \Big\} |0 \rangle .
\end{equation}
The matrix notation here should be understood as a bilinear expression. For example, $\vec{b}^{\dagger} R \bar{\vec{b}}^{\dagger}$ actually means $\sum_{ij}b^\dagger_iR_{ij}\bar{b}_j^\dagger$ where the dagger does not transpose the vector.

The Hamiltonian of the folding picture has the mode expansion in terms of the $b_n$ (with periodic boundary conditions)
\begin{equation}
\begin{aligned}
  H &= \frac{2\pi}{\beta} (L_0 + \bar{L}_0) =  \frac{4\pi}{\beta}  L_0 \\
  &= \frac{4\pi}{\beta}\sum_{\substack{n > 0\\ i=1,2} }  n (b^i_n)^{\dagger} b^i_n \\
  &=  \frac{1}{\pi}(\vec{b}_1^{\dagger} \vec{b}_2^{\dagger} ) (\mathbb{I}_2  \otimes M)
\begin{pmatrix}
  \vec{{b}_1}\\
  \vec{{b}_2}
\end{pmatrix}\\
 &= \frac{1}{\pi} \vec{b}^{\dagger}  (\mathbb{I}_2  \otimes M)  \vec{b} ,
\end{aligned}
\end{equation}
where $L_0+ \bar{L}_0$ are the dilation operator in CFT and we have used the condition $L_0 = \bar{L}_0$ when restricted to the space of the boundary states. The infinite dimensional matrix $M$ is
\begin{equation}
M =  \frac{4\pi^2}{\beta} \text{diag}( 1, 2, \cdots ).
\end{equation}
The partition function in Eq.~\eqref{eq:partition_fun} becomes
\begin{equation}
\label{eq:Zab-bd}
\begin{aligned}
Z_{ab} =& \langle b | e^{-\pi H} | a \rangle \\
=& \langle 0 | \exp\Big\{  \vec{b} R_b( \theta )    \vec{\bar{b}} \Big\}  \exp\Big\{ -\vec{b}^{\dagger}  (\mathbb{I}_2  \otimes M)  \vec{b} \Big\} \\
&\exp\Big\{  \vec{b}^{\dagger} R_a( \theta )    \vec{\bar{b}}^{\dagger} \Big\} | 0\rangle .
\end{aligned}
\end{equation}

In App.~\ref{app:lambda_12}, we obtained the leading order term in the free energy associated with Eq.~\eqref{eq:Zab-bd}. This expression is also obtained by an alternative Casimir energy calculation in App.~\ref{app:gnd_dn_lambda} for one set of the boundary conditions. A na\"ive application of the result however will lead to an apparent contradiction. One notable example is that when $a = b  = {\rm P}$, the free energy given by App.~\ref{app:lambda_12} is $- \frac{1}{12}\beta$, which should actually be {\it zero} because this is the (regularized) free energy on a plane without any interface. Physically this corresponds to the situation that the boundary condition does not change after joining the two chains. Hence the Loschmidt echo will stay at $1$ and the free energy is $0$. This motivates a shift to the free energy
\begin{equation}
\mathcal{F} = - \ln Z_{ab} ( \beta ) + \frac{1}{12} \beta ,
\end{equation}
where $\frac{1}{12}\beta$ is the value of $ \ln Z_{ab} ( \beta )$ when $a = b = {\rm P}$. A more careful inspection in App.~\ref{app:F_correction} shows the origin of the shift: part of it comes from the outer semi-circles in the middle panel of Fig.~\ref{fig:fidel-map} and Fig.~\ref{fig:H-tau_fold}, and another part comes from the non-homogeneous term in the conformal transformation of the stress tensor from annulus to cylinder. 

After incorporating this shift, for the process ($c$ is assumed to be the same as $a$)
\begin{equation}
\label{eq:S_i_S_j}
S_i( \theta_1 ) \rightarrow S_j( \theta_2 ) ,
\end{equation}
the free energy is
\begin{equation}
\mathcal{F}( \beta )  = 
\left\lbrace
\begin{aligned}
  &\frac{1}{2}(|x| - x^2 )\beta  \quad &i = j \\
  &\frac{1}{16}\beta   \quad &i \ne j ,  \\
\end{aligned} \right. \quad x = \frac{\theta_2 - \theta_1}{\pi} .
\end{equation}
We can then set $\beta = 2 \ln L$ and $ 4 \ln t$ (after analytic continuing to real time) to get the fidelity and echo exponent. 

As analyzed in Sec.~\ref{sec:analytic_numerics}, $S_2( \theta)$ interpolates between DD and NN, and $S_1( \theta )$ interpolates between DN and ND. In the region accessible to the numerical calculation in the lattice model, we choose the process ${\rm DD} \rightarrow  \lambda$ to verify
\begin{equation}
\label{eq:result_DDDD}
\mathcal{F} = 
\left\lbrace
\begin{aligned}
\frac{1}{8}\ln L  &\quad\text{fidelity}  \\
\frac{1}{4}\ln t   &\quad \text{echo} .  \\
\end{aligned} \right.  
\end{equation}
The same results have already been obtained for $\lambda = {\rm P}$\cite{stephan_logarithmic_2013,stephan_local_2011,vasseur_universal_2014,vasseur_crossover_2013,kennes_universal_2014}. Another process $\rm{DN} \rightarrow \lambda$ {\iffalse \color{red} in Eq.~\eqref{eq:DNDN}\fi} is used to verify
\begin{equation}
\label{eq:result_DNDN}
\mathcal{F} = 
\left\lbrace
\begin{aligned}
 (x - x^2 )\ln L   &  \quad {\rm fidelity} \\
 2(x - x^2 ) \ln t  & \quad \text{ Loschmidt echo} , \\
\end{aligned} \right. 
\end{equation}
where $\lambda = \tan \theta$ and $x = \frac{\theta}{\pi}$. 

We also use a more artificial process ${\rm P} \rightarrow \lambda$ to check the shift of the curve
\begin{equation}
\label{eq:periodic-case}
\mathcal{F} = 
\left\lbrace
\begin{aligned}
  \Big(|x-\frac{1}{4}| - (x-\frac{1}{4})^2 \Big)\ln L   &  \quad {\rm fidelity} \\
  2\Big(|x-\frac{1}{4}| - (x-\frac{1}{4})^2 \Big) \ln t  & \quad \text{ Loschmidt echo} .\\
\end{aligned} \right. 
\end{equation}



\subsection{Numerical Results and Comparison}
\begin{figure}[h]
\includegraphics[width=1\columnwidth]{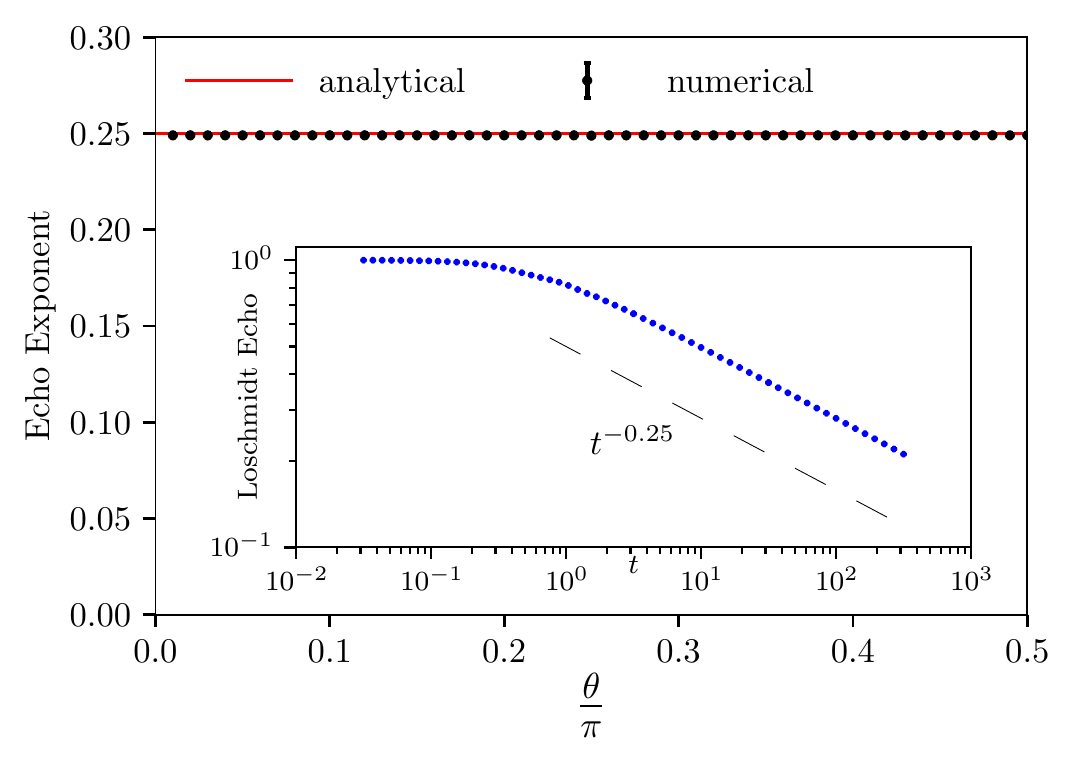}
\caption{The Loschmidt echo decay exponent of the process in $\text{DD}\rightarrow \lambda$ {\iffalse \color{red}Eq.~\eqref{eq:DDDD}\fi}, with gluing condition $S_1(\theta)$. We work with the total system size $N = 30000$ sites, and parameters $m = 10^{-8}$, $k = 1$. The lattice constant is set to unity. The blue dots representing the numerical results lie on the red analytic line. As predicted, the echo exponents are all equal for different values of $\theta$. Inset: An example of Loschmidt echo with $\theta = 0.02 \pi$ shown in log-log scale. The dashline denotes the expected power law of $t^{-0.25}$. Finite size effect does not emerge before $t=10^{3}$, which sets the right boundary of the range we fit. See main text for the curve fitting method.}
\label{fig:DDDD}
\end{figure}

We use the lattice model introduced in Sec.~\ref{sec_sub:free_boson_lattice} to check the analytic results. Our numerical calculations are based on a boson Bogoliubov transformation and the explicit form of the groundstates. The readers are referred to App.~\ref{app:comp_fid_echo} for the technical details. In all the figures, we present the coefficients of the logarithmic terms $\frac{\mathcal{F}}{\ln L}$ and $\frac{\mathcal{F} }{\ln t}$ and call them fidelity and echo exponents respectively. 

We first consider the process ${\rm DD} \rightarrow \lambda$
and show its Loschmidt echo of system size 30000 sites in Fig.~\ref{fig:DDDD}. The inset is a typical Loschmidt echo diagram, whose linearly decreasing behavior in the log-log scale indicates the expected power law decay. We also provide the analytic prediction $\mathcal{L}(t)\sim t^{-0.25}$ (\cf Eq.~\eqref{eq:result_DDDD} as contrast). The exponent (negative of the slope of the line in the log-log plot) is calculated by fitting such diagrams for $\theta = 0.01n \pi$, $n = 1,...,50 $. The fitting is performed before the finite size revival surges and error is estimated by assuming independent and identical Gaussian distribution for each point. We see that the exponents all match with the $\frac{1}{4}$ theoretical line within error. 

We also calculated the companion process ${\rm NN} \rightarrow \lambda$ and obtain identical exponents as in Fig.~\ref{fig:DDDD}. We avoid the technical subtlety of the zero mode by adding a small mass regulator $m=10^{-8}$. While the short time decay pattern is different from the DD case, the long-time behavior and exponents remain the same for both echo and fidelity. We therefore do not present the result here. 

\begin{figure}
  \centering
\includegraphics[width=1\columnwidth]{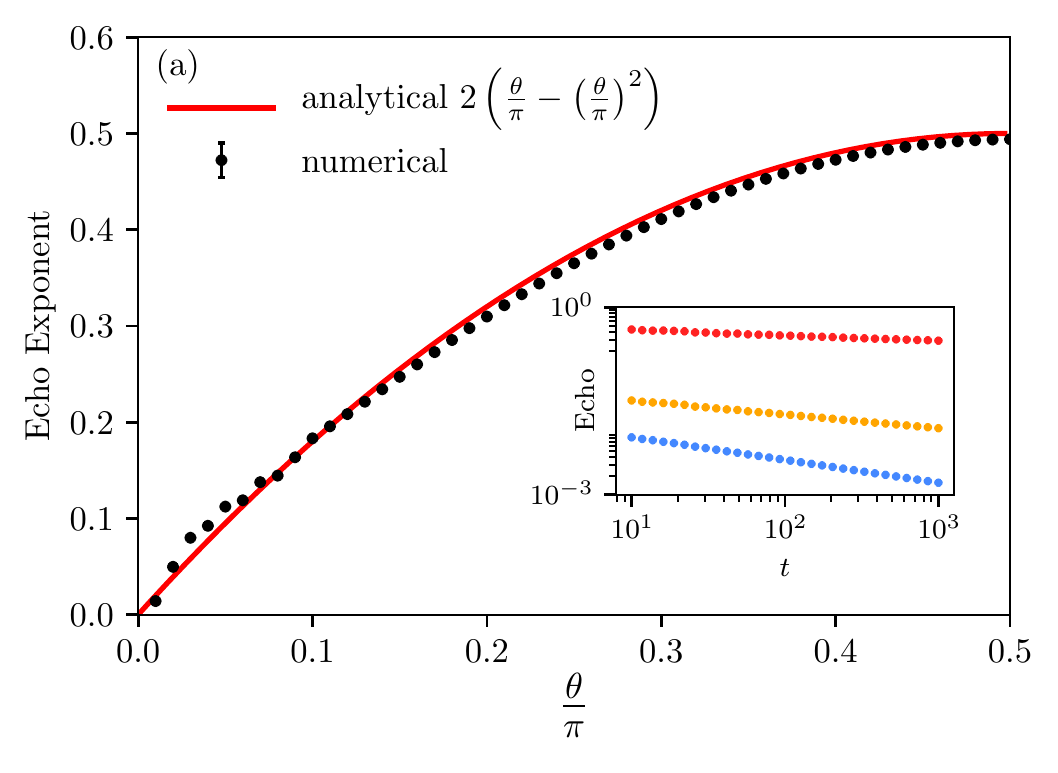}
\includegraphics[width=1\columnwidth]{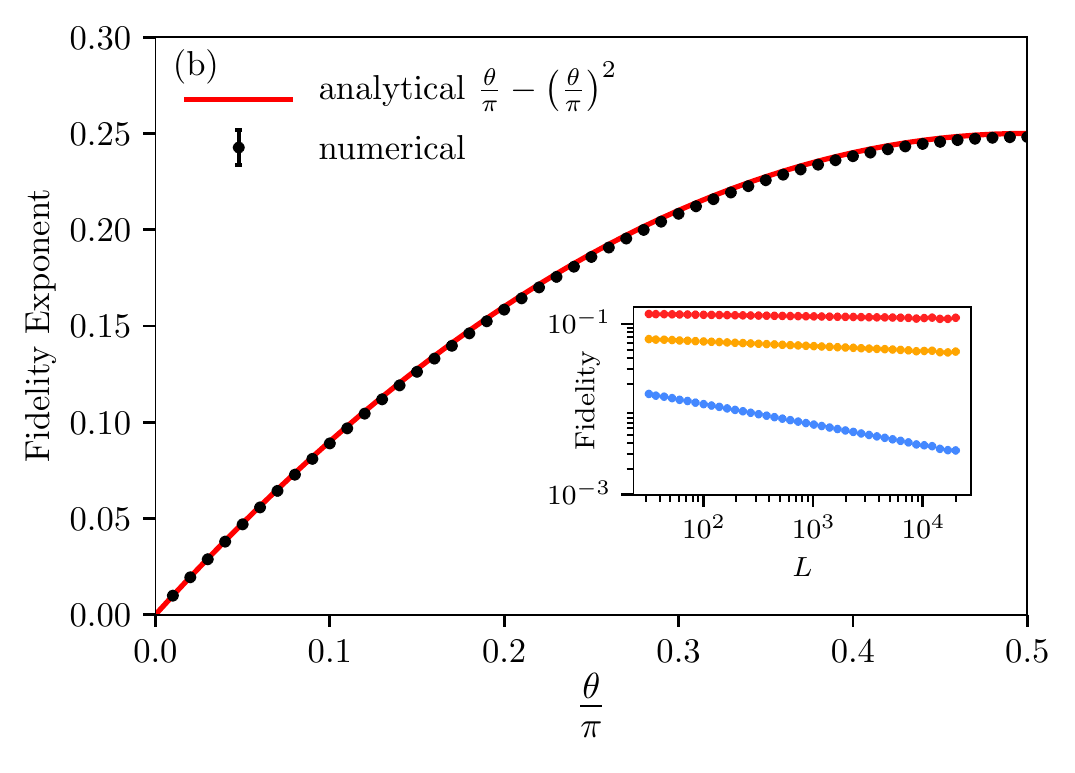}
    \caption{The slope of the free energy for (a) the Loschmidt echo and (b) the bipartite fidelity of the process $\text{DN} \rightarrow \lambda$. {\iffalse in Eq.~\eqref{eq:DNDN}\fi} The total system size is $N=35000$ sites with the same parameters as in Fig.~\ref{fig:DDDD}. The numerical value of exponents follow a quadratic relation as predicted. There is still visible deviation from the analytic results in (a) due to the subtlety of zero mode, see the discussion in the main text. Inset in (a): From the top to bottom, we show the power law decay of the Loschmidt echo with $\theta=0.02\pi, 0.12\pi $ and $0.24\pi$. Finite size effect does not emerge before $t=10^{4}$. We use the same curve fitting method as described in Fig.~\ref{fig:DDDD}.}
      \label{fig:DDNN}
\end{figure}

Next, we analyze the more interesting $\theta$ dependent process ${\rm DN} \rightarrow \lambda$
in which the boundary condition after joining is determined by $S_1(\theta)$. We worked with a system containing $35000$ sites. A direct calculation with the mass regulator does not perform very well in the small $\theta$ regime: the exponent is slightly larger than the theoretical prediction. We therefore turn to another regulator that shift the far end boundary condition DN to $S_1( \delta \theta )$ and consider the following process (see the full notation in Eq.~\eqref{eq:Full_notation_rand()})
\begin{eqnarray}\begin{aligned}
\label{eq:approx_DNDN}
S_1(0)+S_1(\delta\theta)\rightarrow S_1(\theta)+S_1(\delta\theta).
\end{aligned}\end{eqnarray}
Since ${\rm DN} = S_1 (0 )$, taking smaller and smaller $\delta \theta$ should correspond to the original process. This ``shift" regulator works very well for the fidelity calculation, where $\delta \theta = 0.001 \pi$, while moderately good for the Loschmidt echo, where $\delta \theta = 0.003\pi$, see Fig.~\ref{fig:DDNN}. The inset shows the $\theta$-dependence of the power law decay, and the corresponding exponents follow the quadratic relation as predicted in Eq.~\eqref{eq:result_DNDN}. 

\begin{figure}
  \centering
  \includegraphics[width=1\columnwidth]{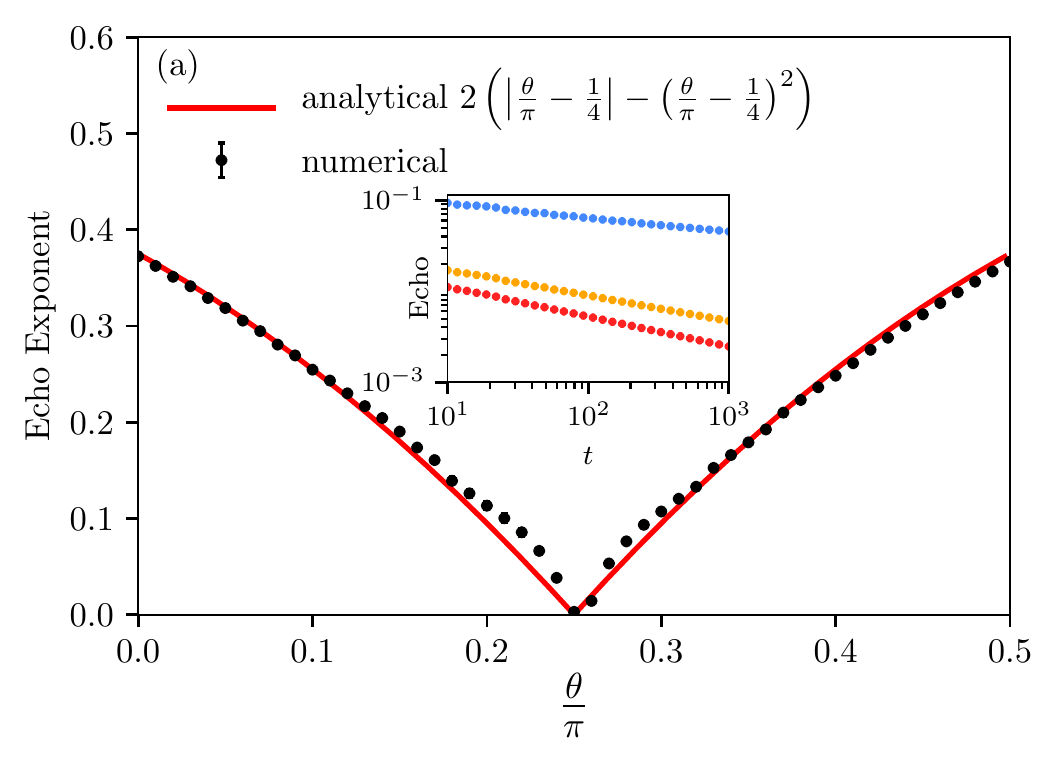}
    \includegraphics[width=1\columnwidth]{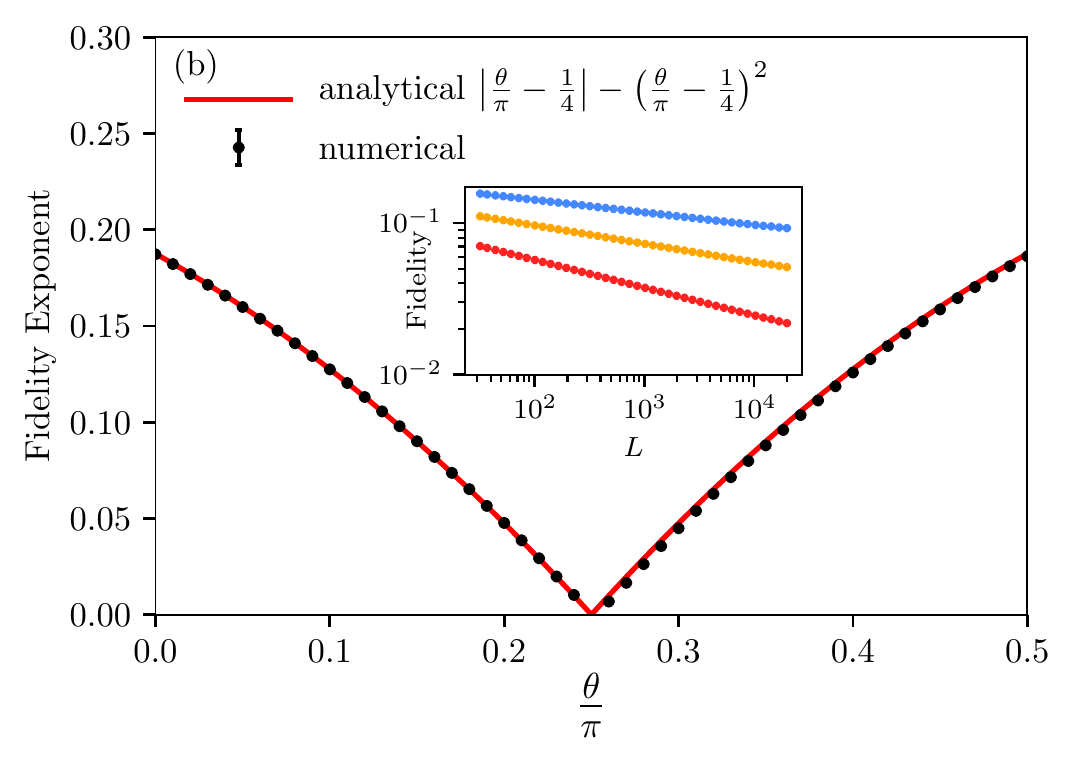}
    \caption{The decay exponent of (a) the Loschmidt echo and (b) the bipartite fidelity of the process $\text{P} \rightarrow \lambda$.{\iffalse in Eq.~\eqref{eq:PP}\fi} The parameters are the same as those in Fig.~\ref{fig:DDNN}. The plot is symmetric with respect to $\theta=0.25\pi$ as predicted. The deviation around $\frac{\pi}{4}$ in (a) is small but visible, see the discussion in the main text. Inset in (a): From top to bottom, we show the power law decay of the Loschmidt echo with $\theta=0.04\pi, 0.08\pi $ and $0.18\pi$. Finite size effect does not emerge before $t=10^{4}$. We use the same curve fitting method as described for Fig.~\ref{fig:DDDD}.}
    \label{fig:PPPP}
\end{figure}

We finally consider the process $\text{P} \rightarrow \lambda$
in Fig.~\ref{fig:PPPP}. Since ${\rm P} = S_1( \frac{\pi}{4} ) $, the zero mode now occurs at $\theta = \frac{\pi}{4}$. We therefore apply the shift regulator there 
\begin{equation}
\begin{aligned}
\label{eq:approx_PPPP}
S_2\left(\frac{\pi}{4}\right)+S_2\left(\frac{\pi}{4}+\delta\theta\right)\rightarrow S_2(\theta)+S_2\left(\frac{\pi}{4}+\delta\theta\right),
\end{aligned}
\end{equation}
where $\delta\theta=0.003\pi$. The $\theta$ dependent exponents are now symmetric about $\theta=\frac{\pi}{4}$ and quadratic on each side, in accordance with Eq.~\eqref{eq:periodic-case}.

Finally, we also provide the data for the process
\begin{equation}
\label{eq:DNP_rand()}
{\rm DN} + {\rm P} \rightarrow \lambda  + {\rm P},
\end{equation}
to evaluate the influence of the boundary condition $c$. It {\rm does} influence the scaling dimension, which is not captured by our analytic calculation. We note that the exponent still follows the quadratic relation with a deficit of $\frac{1}{8}$ at large value of $\theta$. The deficit approaches zero at $\theta=0$ where the setups are the same before and after quench.
\begin{figure}[htb]
\centering
\includegraphics[width=1\columnwidth]{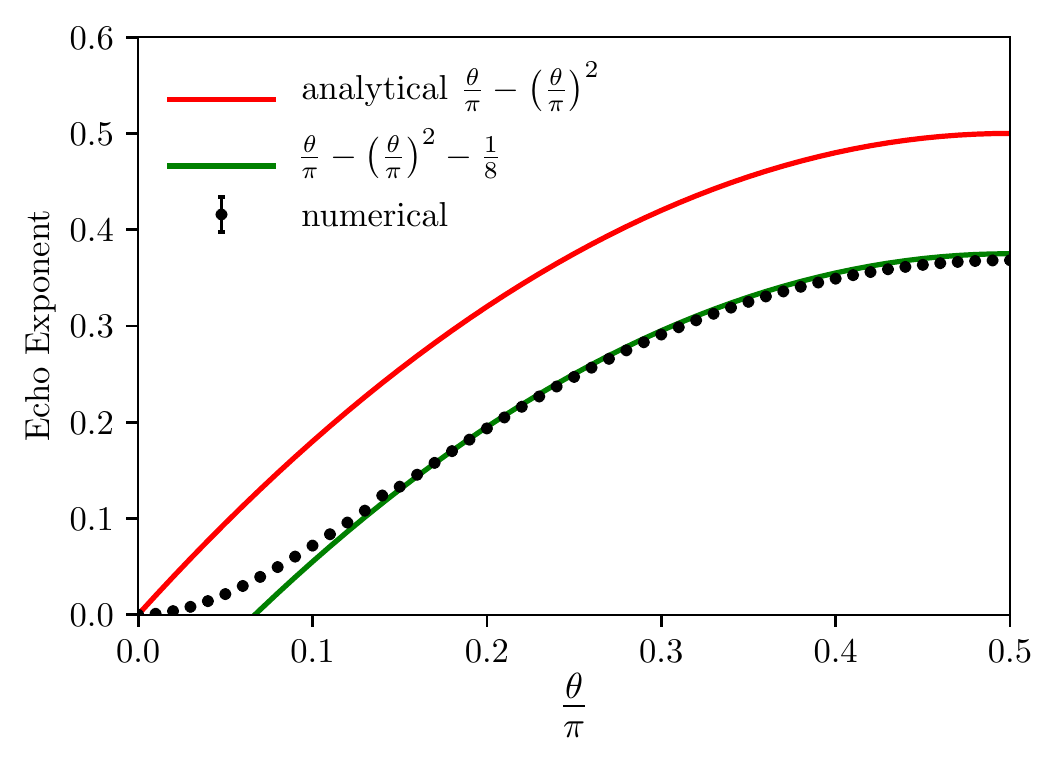}
\caption{The decay exponent of the Loschmidt echo for the process in Eq.~\eqref{eq:DNP_rand()}. The boundary condition $c$, which is now different from $a$, changes the scaling completely from the analytic quadratic curve. The deficit to the analytic value is roughly $\frac{1}{8}$ for $\theta > \frac{\pi}{4}$, It becomes smaller and approaches zero for small $\theta$.}
\label{fig:PDN_fit}
\end{figure}



\section{Discussion}
\label{sec:disc}


In the computations we have done, the results of the fidelity can be converted to that of the Loschmidt echo by the replacement recipe $ \ln L \rightarrow 2 \ln \tau$. The numerical factor of $2$ comes from the fact that the Loschmidt echo has two slit tips. Other than that, we see that they probe the same finite size effect of the free energy associated with the new interfaces. The computation of the (simpler) fidelity is diagnostic and so in the following we will mainly discuss the echo properties. 

In Sec.~\ref{sec_sub:analy_eval}, we have presented the analytic results for the general process $S_i( \theta_1 ) \rightarrow S_j( \theta_2 )$ {\iffalse {\color{red}described in Eq.~\eqref{eq:S_i_S_j}}\fi} (assuming the far end boundary condition $c$ is the same as prior-quench condition $a$). 

We find that if the conformal interfaces are of different types, i.e. $i \ne j$, the (long time) free energy is always $\frac{1}{4} \ln t$, regardless of the theta angles. The two types of conformal interface do not talk to each other because they are imposed on different fields. If we treat $S_1$ as a combination of the Dirichlet and Neumann boundary conditions on the rotated $\phi$ fields as in Eq.~\eqref{eq:rotation_a_basis}, then $S_2$ imposes one of them on the dual field of $\phi$. In the derivation of the $M$ matrix, these two correspond to the parts of the Lorentz group that can not be connected even by taking singular values of $\lambda$. It is then reasonable to find a universal echo between them. The special value of ${\rm DD} \rightarrow P$ also agrees with the existing general CFT result of the completely transparent interface\cite{stephan_logarithmic_2013,stephan_local_2011,vasseur_universal_2014,vasseur_crossover_2013,kennes_universal_2014}. 


For the more interesting case where the boundary conditions are of the same type, we have verified the quadratic angle dependence numerically for ${\rm DN} \rightarrow \lambda$ {\iffalse \color{red}the process in Eq.~\eqref{eq:DNDN}.\fi}. We can first understand the values of several special points on this curve. 
\begin{itemize}
\item $\theta = 0$: This is where the boundary condition does not change before and after the quench, so the Loschmidt echo stays at $1$ and hence the exponent is $0$.
\item $\theta = \frac{\pi}{2}$: This is the process ${\rm DN}\rightarrow {\rm  ND}$. The chain is still disconnected after the change of the boundary conditions. We can thus view the problem as changing the boundary conditions for two independent chains in the left panel of Fig.~\ref{fig:echo}, one from D to N and the other from N to D. The Loschmidt echo can then be viewed as the product of the boundary two point correlation functions of the associated bcc operators $\phi_{\rm DN}$ and $\phi_{\rm ND}$, whose dimensions are both $\Delta = \frac{1}{16}$. From this,
\begin{equation}
\begin{aligned}
  \quad \mathcal{L}(\tau) &\sim |\langle \phi_{{\rm D N}}(0) \phi_{{\rm ND}}(\tau)   \rangle |^2 
|\langle \phi_{{\rm ND}}(0)  \phi_{{\rm DN}}(\tau)   \rangle| ^2\\
& \sim \frac{1}{|\tau|^{8\Delta}} = \frac{1}{|\tau|^{\frac{1}{2}}},
\end{aligned}
\end{equation}
we get the exponent to be $\frac{1}{2}$, which agrees with Fig.~\ref{fig:DDNN}. 
\item $\theta = \frac{\pi}{4}$: This is the process ${\rm DN}\rightarrow {\rm P}$. The exponent $\frac{3}{8}$ agrees with Ref.~\onlinecite{kennes_universal_2014,stephan_logarithmic_2013}, where the difference with the exponent $\frac{1}{4}$ of ${\rm DD} \rightarrow {\rm P}$ is interpreted as twice the dimension of the bcc operator $\phi_{\rm DN}$ ($\Delta = \frac{1}{16}$) that transforms D to N.
\end{itemize}
In general, the result gives the full spectrum of operator dimensions in the ${\rm DN} \rightarrow \lambda $ transition. In the bosonic CFT we consider, the primary fields are the vertex operators $\exp( \nu \phi )$. Depending on the convention, its dimension is a numerical constant times $\frac{\nu^2}{8\pi}$. So if $\nu$ depends linearly on $\theta$ (or $x$), then we will end up with a quadratic relation whose expression can already be fixed by the three special points above. In a rational CFT theory, the number of primary fields is finite. It requires further exploration to identify these bcc operators with the existing primary fields and their physical significance. 

On the other hand, in our lattice boson model, $\theta$ parameterizes the bond interaction between the boundary sites of the chains. Consequently it characterizes the strength of the local perturbation to the Hamiltonian: smaller $\theta$ means smaller change of the bond interaction matrix $\Sigma$ thus a smaller perturbation and vise versa. Therefore, we expect that a larger perturbation will result in a faster decay of the Loschmidt echo, which is reflected by the monotonically increasing decay exponent (the absolute value of the exponent) in the figures.

Our numerical study also shows that the far end boundary condition $c$, which in the large system size limit should not impact the system, {\it does change} the scaling dimension in a way that is not captured by our analytic computation. The reason is that the boundary condition on the far end may introduce additional bcc operators and thus change the free energy. It would be interesting to have a CFT calculation that reproduces the better numerical result in Fig.~\ref{fig:PDN_fit} for the process in Eq.~\eqref{eq:DNP_rand()}. 


This set of the boundary conditions can be realized by connecting two compact bosons. There are already numerous theoretical and experimental works on the boundary conditions of a Luttinger liquid\cite{schmeltzer_zero_1999,anfuso_luttinger_2003,voit_bounded_2000,fabrizio_interacting_1995,egger_applying_1998}, which is the universal compact boson theory of the (bosonized) one-dimensional electron gas\cite{giamarchi_quantum_2015}. For example, gate voltage \cite{egger_applying_1998} may be used to twist the left and right modes of the boson to create a boundary condition interpolating between the normal open and fixed boundary conditions. The interface studied in this paper is a generalization which (in the folding picture) twists the two independent bosons (two left modes plus two right modes) on their connecting ends. An X-ray edge singularity experiment in a quantum wire system, which uses ions to switch on and off the boson interfaces should be plausible to detect the exponents found in this paper.



\section{Conclusion}
\label{sec:conclusion}

In this paper, we analyzed a class of boson conformal interfaces by computing the Loschmidt echo and the bipartite fidelity. 

We began by classifying the boundary states by two types of $S$ matrices $S_1(\theta)$ and $S_2(\theta)$, where the parameter $\theta$ -- the scattering angle -- is determined by the transmission coefficient of the interface. The conventional `DD', `NN' boundary conditions are among the special choices of $\theta$ in $S_1$, and `DN' and P are among the special choices of $S_2$. Generic value of $\theta$ then interpolates between those conventional boundary conditions. A harmonic chain model allows us to realize part of these partially-transmitive boundary conditions in a concrete lattice setting. 

The dynamical behavior of the Loschmidt echo reflects the change of the conformal interfaces during the process described in Eq.~\eqref{eq:S_i_S_j}.
Its power law decay exponent is related to the scaling dimension of the bcc operator that mediates the interfaces. Analytic computation shows that the exponent is always $\frac{1}{4}$ when the change of boundary conditions is made between different types of $S$ matrices ($i \ne j$), regardless of the choice of $\theta$. On the other hand, the exponent depends on the difference of angles $\theta_1 - \theta_2$ as a quadratic relation when the change is made between the same type of $S$ matrices ($i = j$).

These two features are tested in three typical processes in the numerical calculation of the harmonic chains. After using suitable regulators for the zero-mode problem, the numerical results agree with the analytic calculation within error. Although tangential to the non-equilibrium dynamics, the fidelity calculation is used as a diagnostic tool and shows better agreement of the exponent, providing more confidence about our analytic results. 

We proposed that the Loschmidt echo exponent in principle should be detectable in an X-ray edge singularity type experiment on the quantum wire systems. 



\begin{acknowledgments}
We are grateful for the discussion and suggestions provided by Thomas Faulkner, Xueda Wen, Shinsei Ryu, Michael Stone, Taylor Hughes and Romain Vasseur during the course of this research. We would like to thank Shinsei Ryu, Michael Stone and Taylor Hughes for reading the manuscripts. We also acknowledge the valuable questions and comments from the referees. 
TZ is supported by the National Science Foundation under grant number NSF-DMR-1306011.
M.L. is supported by US NSF under grant DMR 1351895-CAR and US NSF Emerging Frontiers in Research and Innovation grant EFMA-1627184
This work made use of the Illinois Campus Cluster, a computing resource that is operated by the Illinois Campus Cluster Program (ICCP) in conjunction with the National Center for Supercomputing Applications (NCSA) and which is supported by funds from the University of Illinois at Urbana-Champaign.
\end{acknowledgments}

\appendix

\section{General Boundary State Amplitude}
\label{app:lambda_12}

\begin{figure}[h]
\centering
\includegraphics{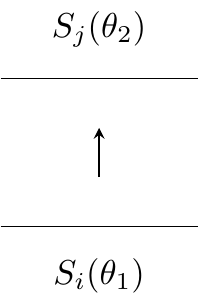}
\caption{Partition function between the two boundary states of $S_i( \theta_1)$ and that of $S_j( \theta_2 )$}
\label{fig:fig_lambda_1_lambda_2}
\end{figure}

In this appendix, we calculate the amplitude between general boundary states defined in Eq.~\eqref{eq:Zab-bd}
\begin{equation}
\begin{aligned}
Z_{ab} &= \langle 0 | \exp\Big\{  \vec{b} R_b( \theta )    \vec{\bar{b}} \Big\}  \exp\Big\{ -\vec{b}^{\dagger}  (\mathbb{I}_2  \otimes M)  \vec{b} \Big\} \\
&\exp\Big\{  \vec{b}^{\dagger} R_a( \theta )    \vec{\bar{b}}^{\dagger} \Big\} | 0\rangle,
\end{aligned}
\end{equation}
where
\begin{equation}
\begin{aligned}
M &=  \frac{4\pi^2}{\beta} \text{diag}( 1, 2, \cdots ), \quad  \mathbb{I}_2 = \text{diag}( 1, 1), \\
R_i &= S_i( \theta ) \otimes \mathbb{I}.
\end{aligned}
\end{equation}
The graphical representation of the partition function is shown in Fig.~\ref{fig:fig_lambda_1_lambda_2}. Using the identity Eq.~\eqref{eq:second_id_in_app_pf_of_id} proven in App.~\ref{app:pf_of_id}, we have
\begin{equation}
Z_{ab} = \frac{1}{\det(1 - R_a^{\dagger} e^{- \mathbb{I}_2 \otimes M} R_b )}
\end{equation}
From $|\det( R_a R_b^{\dagger})|  = 1$, free energy becomes
\begin{equation}
F = - \ln |Z_{ab}| = \ln |\det ( R_a R_b^{\dagger} - e^{- \mathbb{I}_2 \otimes M} )| .
\end{equation}
There are two cases to be considered, and we only take out the leading order term in $\beta$. 
\begin{itemize}
\item {\it case 1: }$S_1( \theta_1 ) \rightarrow S_2 ( \theta_2 ) $, the free energy is
\begin{equation}
\begin{aligned}
F & = \ln |\det ( R_1( \theta_1 )  R_2^{\dagger}( \theta_2 )  - e^{- \mathbb{I}_2 \otimes M} )| \\
  & = \ln \left| \det
\begin{bmatrix}
-\cos 2 \Delta \theta \mathbb{I} - e^{-M}   & -\sin 2 \Delta \theta \mathbb{I}\\
- \sin 2\Delta \theta \mathbb{I}  &   \cos 2 \Delta \theta \mathbb{I} - e^{-M} \\ 
\end{bmatrix} \right| \\
& = \sum_i \ln [ 1 -  e^{- 2 \lambda_i }  ] \\
& = \frac{\beta}{4\pi^2} \int_0^{\infty} dx \ln [ 1 - e^{-2x} ]  = - \frac{1}{48 }\beta .
\end{aligned}
\end{equation}
\item {\it case 2:} $S_i( \theta_1 ) \rightarrow S_i( \theta_2 )$, where $i = 1 $ or $ 2$, 
\begin{equation}
\begin{aligned}
F & = \ln \det 
\begin{bmatrix}
\cos 2 \Delta \theta \mathbb{I} - e^{-M}   & \sin 2 \Delta \theta \mathbb{I}\\
- \sin 2\Delta \theta \mathbb{I}  &   \cos 2 \Delta \theta \mathbb{I} - e^{-M} \\ 
\end{bmatrix} \\
& = \sum_i \ln [ 1 - 2 \cos 2 \Delta \theta e^{- \lambda_i } + e^{- 2 \lambda_i }  ] \\
& = \frac{\beta}{4\pi^2} \int_0^{\infty} dx \ln [ 1 - 2 \cos 2 \Delta \theta e^{-x} + e^{-2x} ] ,
\end{aligned}
\end{equation}
where $\Delta \theta = \theta_2 - \theta_1$. This integral is an even function of $\Delta \theta$ and the $\Delta \theta > 0$ case reduce to the polylog and Bernoulli polynomial
\begin{equation}
\begin{aligned}
  F &= \frac{\beta}{4\pi^2} \left[ - \text{Li}_2 ( e^{2i |\Delta \theta|} ) - \text{Li}_2 ( e^{- 2i |\Delta \theta|} ) \right] \\
  & = \frac{\beta}{4\pi^2}  \left[ - 2\pi^2 B_2 (|x|) \right] \\
  &= - \frac{\beta}{2} B_2( |x| )  = \frac{\beta}{2} (| x| - x^2 - \frac{1}{6} ),
\end{aligned}
\end{equation}
where $x = \frac{\Delta \theta}{ \pi}$. 
\end{itemize}



\section{Alternative Approach to ${\rm DN} \rightarrow \lambda$ Amplitude}
\label{app:gnd_dn_lambda}

\begin{figure}[h]
\centering
\includegraphics[width=\columnwidth]{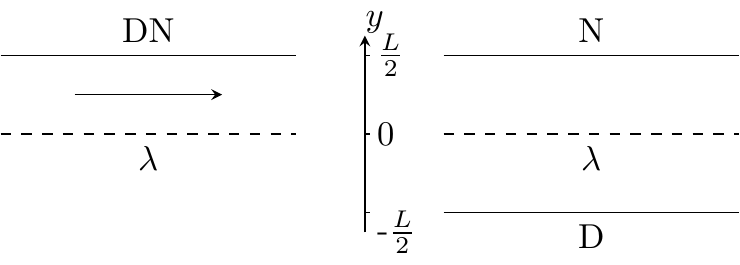}
\caption{Partition function of Hamiltonian with DN and $\lambda$ boundary conditions. We unfold the cylinder and the new stripe has N and D boundary conditions on the top and bottom plus a $\lambda$ junction in the middle. The length $L$ here is the height of the unfolded cylinder $2\pi$. }
\label{fig:Fig_gnd_dn_lambda}
\end{figure}

In this appendix, we calculate the amplitude for the setup shown in Fig.~\ref{fig:Fig_gnd_dn_lambda}. In particular, the unfolded configuration has D/N boundary conditions at $y = \pm \frac{L}{2}$ and conformal interface $\lambda$ at $y = 0$. The general solutions can be written as
\begin{equation}
\label{eq:normalized_f}
f(k, y) = 
\left\lbrace
\begin{aligned}
  A_1 e^{i kt} \cos\left(ky +\frac{1}{2}kL \right) &  \quad y < 0  \\
  A_2 e^{ikt}  \sin\left(ky - \frac{1}{2}kL \right) & \quad y > 0 .  \\
\end{aligned} \right. 
\end{equation}
As demonstrated in Sec.~\ref{sec_sub:general_formulation}, if we denote $f(k,y<0)\equiv\phi_1$ and $f(k,y>0)\equiv\phi_2$, the boundary condition at the junction becomes
\begin{eqnarray}\begin{aligned}
\frac{\partial_x \phi_1}{ \partial_t \phi_1} = \lambda^2 \frac{\partial_x \phi_2}{ \partial_t \phi_2} = \tan^2 \theta\frac{\partial_x \phi_2}{ \partial_t \phi_2}, \quad \theta \in \left[0,\frac{\pi}{2} \right]  ,
\end{aligned}\end{eqnarray}
which implies
\begin{equation}
\label{eq:momentum_gnd_dn_lambda}
k = \frac{2\pi}{L}\left( n \pm \frac{\theta}{\pi} \right),  \quad n\in\mathbb{Z}.
\end{equation}
It is evident that the momentum $k$ is shifted from an integer multiple of $\frac{2\pi}{L}$ due to the $\lambda$ boundary condition in the middle. 

The normalized eigenfunctions in Eq.~\eqref{eq:normalized_f} serves as an orthonormal basis in the mode expansion, we thus have
\begin{equation}
\label{eq:H_in_gnd_dn_lambda}
H = \frac{1}{2} \sum_{n \in \mathbb{Z} } |k|  \left(a^{\dagger}_n a_n + \frac{1}{2} \right) ,
\end{equation}
where the momentum $k$ is defined in Eq.~\eqref{eq:momentum_gnd_dn_lambda}, and the creation and annihilation operators are defined as usual
\begin{equation}
\begin{aligned}
a_n = \frac{1}{\sqrt{2}} \left( \sqrt{ |k|g} \phi_n + \frac{i }{\sqrt{|k|g} }\pi_n  \right) ,\\
a^{\dagger}_n = \frac{1}{\sqrt{2}} \left( \sqrt{ |k|g} \phi_n - \frac{i }{\sqrt{|k|g} }\pi_n  \right) .\\
\end{aligned}
\end{equation}

The Casimir energy is the vacuum energy brought up by the finite size of the setup. From Eq.~\eqref{eq:H_in_gnd_dn_lambda} and using $x\equiv \frac{\theta}{\pi}$ in Eq.~\eqref{eq:momentum_gnd_dn_lambda}, we have
\begin{equation}
\begin{aligned}
E_c &= \frac{1}{4} \sum_{n \in \mathbb{Z}} | k| = \frac{\pi}{2L} \left( \sum_{n \in \mathbb{Z}}  | n + x | + \sum_{n \in \mathbb{Z}}  | n - x |  \right) \\
&= \frac{\pi}{2L} \Bigg( \sum_{n\ge 0 }  ( n + x ) + \sum_{n< 0 }  ( -n - x )  \\
&\quad \quad + \sum_{n \le 0}  (- n + x ) + \sum_{n > 0} ( n- x )  \Bigg) \\
&= \frac{\pi}{2L} \left( 2\sum_{n\ge 0 }  ( n + x ) + 2\sum_{n > 0} ( n- x )  \right).
\end{aligned}
\end{equation}
We use the Hurwitz zeta function
\begin{equation}
\zeta_{\rm H}(s,x) = \sum_{n=0}^\infty\frac{1}{(n+x)^s}
\end{equation}
to regularize the sum, 
\begin{equation}
\begin{aligned}
E_c &= \frac{\pi}{L} \Big[\sum_{n \ge 0 } ( n + x )^{-s} + \sum_{n \ge 0 }  ( n - x)^{-s}  -  (-x)^{-s} \Big]\Big|_{s = -1} \\
&= \frac{\pi}{L} \left[ \zeta_{\rm H}( -1, x ) + \zeta_{\rm H}( -1, -x ) +  x \right] \\
&= \frac{1}{2} \left( - x^2 + x - \frac{1}{6}\right),
\end{aligned}
\end{equation}
where in the last line we use $L = 2\pi$ for the unfolded geometry. 

Thus the free energy in the large $\beta$ limit is
\begin{equation}
F = \beta E_c = - \frac{\beta}{2} B_2( x) ,
\end{equation}
which agrees with the boundary state calculation in Sec.~\ref{app:lambda_12}. 



\section{Corrections to the Free Energy}
\label{app:F_correction}

In the course of deriving the free energy subject to various boundary conditions, we use conformal transformation to convert the spacetime diagram with slits to a cylinder diagram, where the boundary state calculation in App.~\ref{app:lambda_12} (and ground state energy calculation in App.~\ref{app:gnd_dn_lambda}) is applicable. However, the free energy is not invariant under the conformal transformation since the boundaries partially break the conformal symmetry. In this Appendix, we point out two corrections -- one from the outer boundary regulator and the other from the inhomogeneous Schwartzian term to get the correct exponent of the fidelity and Loschmidt echo. 

It is discussed in Cardy and Peschel's work\cite{cardy_finite-size_1988} that the boundary will contribute logarithmic term in the free energy,
\begin{equation}
F = - \frac{c}{6} \left( \int_M  K(x) d^2x + \int_{\partial M} k_g ds \right)  \ln L , 
\end{equation}
where $M$ is a 2d smooth manifold, $K(x)$ is the Gaussian curvature, $k_g$ is the geodesic curvature of the boundary of the manifold and $L$ is the system's characteristic length. 

The boundary term was not previous noticed in the literature, but is actually important even in the simplest example of the disk free energy. Consider an annulus on flat space with inner radius $r_1$ and outer radius $r_2$. Its free energy is
\begin{equation}
F({\rm annulus}) = -  \frac{c}{6} \ln \frac{r_2}{r_1} .
\end{equation}
On the other hand the free energy of a disk of radius $r_2$ is
\begin{equation}
  F( {\rm disk} ) = - \frac{c}{6} \ln \frac{r_2}{a},
\end{equation}
where $a$ is the short distance regulator. The disk free energy is completely contributed by its outer boundary with other parts being conformal invariant. In fact, $K = 0, k_g = \frac{1}{r}$ for disk, and so 
\begin{equation}
F( {\rm disk}  ) = - \frac{c}{6}  \left( \int_{\partial M} k_g ds \right)  \ln \frac{r_2}{a}  = - \frac{c}{6}  \ln \frac{r_2}{a} , 
\end{equation}
where $a$ is the short distance cut-off. 

One can then interpret the annulus free energy as additive contributions from its outer and inner surfaces
\begin{equation}
F( {\rm annulus} ) = -  \frac{c}{6} \ln \frac{r_2}{a} +  \frac{c}{6} \ln \frac{r_1}{a} = - \frac{c}{6} \ln \frac{r_2}{r_1}.
\end{equation}
An annulus becomes a disk when its inner radius is of order $a$, and we can see that the contribution from the inner surface $\frac{c}{6} \ln \frac{r_1}{a}$ becomes negligible compared to the one from the outer surface.  

A similar outer surface logarithmic term also appears in the middle panel of Fig.~\ref{fig:fidel-map}. The conformal map from the $z$ plane to $\xi$ plane bring the strip (with the small blue semi-circle) to the upper half plane with the semi-circle around $z = 1$ extracted. This is in close analogy with the truncated corner calculation in Ref.~\onlinecite{cardy_finite-size_1988}. In order to evaluate this diagram, we manually add the large blue semi-circle as IR cut-off, at the price of introducing an additional contribution  $-\frac{c}{6} \ln \frac{r_2}{a}$ of free energy which should not be there. 

The same thing happened in Fig.~\ref{fig:H-tau_fold} with a slightly different mechanism. In the slit diagram (left panel in Fig.~\ref{fig:H-tau_fold}), the regulators all have radii that are at the order of the short distance cut-off. They will have negligible contributions to the free energy. However, in the new $\xi$ plane, we implicitly switch to a {\it new short distance regulator} such that only the blue semi-circle around $0$ contributes negligibly. The outer surface radius, despite being the image of a small semi-circle on the $z$ plane, will contribute a $-\frac{c}{6} \ln \frac{r_2}{a}$ term on the $\xi$ plane that should not be there. 

Therefore in both cases we should compensate $\frac{c}{6} \ln \frac{r_2}{a}$. Using the cylinder parameters in App.~\ref{app:lambda_12}, the $\xi$ plane and $z$ plane free energy are related through
\begin{equation}
F_{z} = F_{\xi} + \frac{c}{6} \beta .
\end{equation}
for both the fidelity and Loschmidt echo. 

The annulus on the $\xi$ plane is called the staircase geometry in Ref.~\onlinecite{cardy_finite-size_1988} due to its evolution in angular direction. The traditional radial quantization however has radial direction to be the time. One can show that the Hamiltonian of the staircase and rectangle has a shift due to the Schwartzian\cite{cardy_finite-size_1988} of the conformal transform
\begin{equation}
H_{\xi} = H_{w} - \frac{c}{24\pi} \beta .
\end{equation}
After the evolution for $2\pi$ (in the folding picture, the evolution is only $\pi$ but there are two bosons), the difference in the free energy is
\begin{equation}
F_{\xi} = F_{w} - \frac{c}{12} \beta .
\end{equation}
Gathering the two terms, we obtain the missing correction $\frac{c}{12} \beta$ between the slit and cylinder diagram, 
\begin{equation}
F_{z} = F_w + \frac{c}{12}\beta.
\end{equation}



\section{Winding Modes of Compact Bosons}
\label{app:compact_diff_boson}

In this appendix, we address the issue of the winding modes of the compact bosons. In the main text, we have exclusively worked with the oscillator modes of the free bosons. Here, we shall show that winding modes for the compactified bosons will have no contribution to the fidelity or Loschmidt echo in the leading order. Therefore, our results are ready to be applied in the case where two compactified bosons of different radii are connected by a conformal interface\cite{PhysRevLett.118.136801}. 

Our derivation follows the general multi-component boson constraints in Ref.~\onlinecite{affleck_quantum_2001,oshikawa_boundary_2010}. A review of the detailed parameterization of the states can be found in Ref.~\onlinecite{sakai_entanglement_2008}. 

\subsection{Mode Expansion of Compact Boson}
Suppose the boson is compactified as $\phi =  \phi + 2\pi R$, using the notation in Ref.~\onlinecite{di_francesco_conformal_1997}, we have the following mode expansion 
\begin{widetext}
\begin{equation}
\label{eq:boson-mode-exp}
\begin{aligned}
\phi( z, \bar{z}) = &\phi_0 -i \left( \frac{n}{4\pi g  R} + \frac{m R }{2} \right)  \ln z + \frac{i}{\sqrt{4\pi g}} \sum_{n\ne 0 } \frac{a_n}{n} z^{-n } -i \left( \frac{n}{4\pi g R} - \frac{m R }{2} \right)  \ln \bar{z} + \frac{i}{\sqrt{4\pi g}} \sum_{n\ne 0 } \frac{\bar{a}_n}{n} \bar{z}^{-n } ,
\end{aligned}
\end{equation}\end{widetext}
where $n,m$ are the momentum and winding modes quantum numbers. 

In Ref.~\onlinecite{di_francesco_conformal_1997}, the quantization is performed on an equal time space. The boundary state we need here however lives on $x = 0$ -- an equal-space slice. We therefore compact the theory in the time direction with period $T$, and identify the holomorphic and anti-holomorphic coordinates as 
\begin{equation}
\label{eq:zzbar}
z = \exp( 2 \pi i \frac{t - x}{L}), \qquad \bar{z} = \exp( 2 \pi i \frac{t + x}{L}).
\end{equation}
This corresponds to exchange the $x$ and $t$ in Ref.~\onlinecite{di_francesco_conformal_1997}. 

We further identify
\begin{equation}
\begin{aligned}
  a_0 &= \sqrt{ 4 \pi g } \left( \frac{n}{4\pi g R} + \frac{m R }{2} \right), \\
   \bar{a}_0 &= \sqrt{ 4 \pi g } \left( \frac{n}{4\pi g R} - \frac{m R }{2} \right),
  \end{aligned}
\end{equation}
to obtain the expression uniform for all the modes
\begin{equation}
\begin{aligned}
i \partial_z \phi =  \sum_n \frac{1}{\sqrt{4\pi g}} a_n z^{-n-1} .
\end{aligned}
\end{equation}

\subsection{Gluing Condition for the Winding Modes}
\label{app_sub:compact_gluing_boundary}

We recall that the gluing condition is written as
\begin{equation}
\begin{aligned}
\label{eq:def_S_in_app}
\begin{pmatrix}
\partial_+\phi^1\\
\partial_-\phi^2
\end{pmatrix}
=S(\theta)
\begin{pmatrix}
\partial_-\phi^1\\
\partial_+\phi^2
\end{pmatrix}.
\end{aligned}
\end{equation}
Upon folding $\phi^2$ to the negative axis, its $\partial_x$ derivative becomes $-\partial_x$, so 
\begin{equation}
\begin{bmatrix}
\partial_{+} \phi^1 \\
\partial_{+} \phi^2 \\
\end{bmatrix}
 = S
\begin{bmatrix}
\partial_{-} \phi^1 \\
\partial_{-} \phi^2 \\
\end{bmatrix}.
\end{equation}
In terms of the holomorphic coordinates defined in Eq.~\eqref{eq:zzbar}, 
\begin{equation}
\partial_{+} = \frac{4\pi i }{T} \bar{z} \partial_{\bar{z}} ,\qquad \partial_{-} = -\frac{4\pi i }{T} z\partial_{z},
\end{equation}
the $S$ matrix establishes a relation between the modes
\begin{equation}
\begin{aligned}
\label{eq:def_S_in_app_2}
\sum_{n\in\mathbb{Z}}
\begin{pmatrix}
\bar{a}_n^1\bar{z}^{-n}\\
\bar{a}_n^2\bar{z}^{-n}
\end{pmatrix}
= -S(\theta)
\sum_{n\in\mathbb{Z}}
\begin{pmatrix}
a_n^1{z}^{-n}\\
a_n^2{z}^{-n}
\end{pmatrix}.
\end{aligned}
\end{equation}
At the boundary $x = 0$, $\bar{z}=z^{-1}$, we have
\begin{equation}
\begin{aligned}
a^i_n + (S^{-1}_{ij})\bar{a}^j_{-n}=0.
\end{aligned}
\end{equation}
The solution of the $n \ne 0$ constraints is exactly the boundary state in Eq.~\eqref{eq:bd_state}. 

We specialize to $S = S_1$ to solve the $ n = 0$ constraint. We introduce the compactification lattice and its dual\cite{affleck_quantum_2001,oshikawa_boundary_2010}
\begin{equation}
\label{eq:lattice}
\vec{M} = (m_1 2 \pi R_1, m_2 2\pi  R_2)^\top, \quad  \vec{M}^* = (\frac{n_1}{R_1}, \frac{n_2}{R_2})^\top,
\end{equation}
to rewrite the zero mode part as 
\begin{equation}
  a_0^i + S^{-1} _{ij} \bar{a}_{0}^j = 0 \,\, \implies \,\, ( \vec{M} + \frac{1}{g}\vec{M}^* ) = S_1 ( -\vec{M} + \frac{1}{g}\vec{M}^* ),
\end{equation}
which is basically the multi-component boson winding constraints given in \onlinecite{affleck_quantum_2001,oshikawa_boundary_2010}. The solution gives the interface parameter $\lambda$
\begin{equation}
\begin{aligned}
\label{eq:S_1_constraint}
\lambda = \tan\theta=\frac{n_2R_1}{n_1R_2}=-\frac{m_1R_1}{m_2R_2},
\end{aligned}
\end{equation}
and the conformal boundary state
\begin{equation}\begin{aligned}
\label{eq:S1bd-state}
g_{S_1}\sum_{S_1}e^{in_1\phi_0-im_1\bar{\phi}_0}|n_1,m_1\rangle|n_2,m_2\rangle,
\end{aligned}\end{equation}
where $\sum_{S_1}$ is the summation consistent with the constraint in Eq.~\eqref{eq:S_1_constraint}. The $g$-factor can only be determined by the Cardy condition\cite{cardy_boundary_2004}. Since it is not important for what follows, we shall not include the calculation here. 

Since $S = S_2$ is effectively $S_1$ on the dual boson, we can expect that it will end up in the same expression as in Eq.~\eqref{eq:S1bd-state}, but with a different constraint on the winding number
\begin{equation}
\vec{M} = \frac{\cot \theta}{g} 
\begin{bmatrix}
0 & -1\\
1 & 0 \\                                
\end{bmatrix}
\vec{M}^*.
\end{equation}

\subsection{Winding Mode Contribution to the Partition Function}
\label{app_sub:winding_contribution}
We now calculate the winding mode part of the partition function as shown in Fig.~\ref{fig:fig_lambda_1_lambda_2}
\begin{equation}\begin{aligned}
Z=\langle S_j( \theta_2 )|e^{-\pi H}|S_i(\theta_1 )\rangle,\qquad H=\frac{2\pi}{\beta}(L_0+\bar{L}_0).
\end{aligned}\end{equation}
For boundary states, we can simply replace $L_0$($\bar{L}_0$) with $a_0$($\bar{a}_0$). 

For the amplitude between the same boundary states $\lambda_1=\lambda_2$, we have the winding mode contribution as
\begin{equation}
\begin{aligned}
\label{eq:Z0}
Z_0 \le  \sum_{S_i} g_{\,\!_{S_i} }g_{\,\!_{S_j} } \exp\Big\{- \frac{4\pi}{\beta} 2 \pi g ( \frac{n_1}{ 4 \pi R_1 g} + \frac{m_1 R_1 }{ 2} )^2 \Big\},
\end{aligned}
\end{equation}
where the equality only is only taken when the two boundary states are identical. 

In the limit $\beta\rightarrow\infty$, Eq.~\eqref{eq:Z0} can be approximated by a simple two dimensional integral
\begin{equation}\begin{aligned}
Z_0\approx g_{\,\!_{S_i} }g_{\,\!_{S_j} }\beta\int dxdy\exp\Big\{-8 \pi^2 g ( \frac{x}{ 4 \pi R_1 g} + \frac{y R_1 }{ 2} )^2 \Big\}.
\end{aligned}\end{equation}
The winding mode thus can contribute at most a $\ln\beta$ term to the free energy. Compared to the result in App.~\ref{app:lambda_12}-\ref{app:gnd_dn_lambda}, we conclude that the winding mode contribution will not present in the leading order of the large $\beta$ limit.



\section{Conformal Interface in Free Bosonic Lattice}
\label{app:interface_free_boson}

In this appendix, we demonstrate how to realize the conformal interface in a lattice harmonic chain defined in Sec.~\ref{sec_sub:free_boson_lattice},
\begin{equation}
\begin{aligned}
H =& \frac{1}{2} \sum_i \pi_i^2  +  \frac{1}{2} \sum_{i\ne 0 }  ( \phi_i - \phi_{i+1} )^2  \\
 &+ \frac{1}{2} \begin{pmatrix}  \phi_0, \phi_1 \end{pmatrix}
\begin{bmatrix}
1 + \Sigma_{11}  & \Sigma_{12} \\
\Sigma_{21} &  1 + \Sigma_{22} \\
\end{bmatrix}
\begin{pmatrix}
  \phi_0 \\
  \phi_1 
\end{pmatrix},
\end{aligned}
\end{equation}
where the matrix $\Sigma$ parameterizes the two-site interaction between site 0 and 1. We set up the plane wave scattering problem across the interface with the following ansatz (the use of $(n-1)$ in $\phi_n^B$ simplifies the calculation)
\begin{equation}
\label{eq:ansatz}
\phi_n
= \left\lbrace
  \begin{aligned}
	& A_{-} e^{i \omega t  - inka}  + A_{+} e^{i \omega t  + inka}  & \quad  n \le 0 \\
	& B_{-} e^{i \omega t  - i(n-1)ka}  + B_{+} e^{i \omega t  + i(n-1)ka} & \quad n \ge 1 ,\\
  \end{aligned} \right. 
 \quad 
\end{equation}
where $a$ is the lattice constant. The solution on both semi-infinite chains are gapless with the dispersion relation $\omega = \left|2\sin\frac{ka}{2}\right|$. The $S$ matrix connecting them can be found by relating the incoming and outgoing amplitudes

\begin{equation}
\label{eq:discrete_S}
\begin{aligned}
\begin{pmatrix}
A_{+} \\
B_{-}\\
\end{pmatrix}
=&-
\begin{bmatrix}
\Sigma_{11} +e^{ika} & \Sigma_{12}\\
\Sigma_{21} & \Sigma_{22} + e^{ika}
\end{bmatrix}^{-1} \\
&\begin{bmatrix}
\Sigma_{11} + e^{-ika} & \Sigma_{12} \\
\Sigma_{21} & \Sigma_{22}  + e^{-ika}  \\
\end{bmatrix}
\begin{pmatrix}
A_{-}\\
B_{+}\\
\end{pmatrix},
\end{aligned}
\end{equation}
and the explicit expression is
\begin{widetext}
\begin{equation}
  S = \frac{-1}{ \det \Sigma  + e^{-ika} \text{tr} \Sigma   + e^{-2ika}}
\begin{bmatrix}
\det \Sigma+ \Sigma_{11} e^{-ika} + \Sigma_{22} e^{ika}+1  & -2i \sin ka \Sigma_{12}  \\
-2i \sin ka \Sigma_{21} &  \det \Sigma+ \Sigma_{11} e^{ika} + \Sigma_{22} e^{-ika}+1\\
\end{bmatrix}.
\end{equation}
\end{widetext}

The reflection and transmission coefficients associated with this interface contained in the $S$ matrix and both of them have to be $k-$independent to form a conformal interface\cite{peschel_exact_2012}. A necessary condition is that $|S_{12}|$ must be $k$-independent. If $\Sigma_{12} = \Sigma_{21} = 0 $, we have 
\begin{equation}
  S_{11} = - \frac{\Sigma_{11} + e^{ika} }{\Sigma_{11} + e^{-ika}},
\end{equation}
which is not scale invariant. The only remaining possibility is
\begin{equation}
\label{eq:Sigma_condition}
\det \Sigma = -1, \, \, \text{tr} \Sigma = 0,
\end{equation}
which leads to a scale invariant S-matrix 
\begin{equation}
S = \frac{1}{1 - e^{-2ika } } ( -2i \sin ka ) \Sigma
 = - e^{ika} \Sigma.
\end{equation}
In this continuum limit where $a\rightarrow0$, the matrix $\Sigma$ can be parameterized as
\begin{equation}
\Sigma = -\lim_{a \rightarrow 0 } S = 
\begin{bmatrix}
\frac{\lambda^2- 1}{1 + \lambda^2} & \frac{-2\lambda }{1 + \lambda^2} \\
\frac{-2\lambda }{1 + \lambda^2} & \frac{1- \lambda^2}{1 + \lambda^2} \\
\end{bmatrix},
\end{equation}
where $\lambda\in\mathbb{R}$ is the parameter for $S_1(\theta)$, as introduced in Sec.~\ref{sec:notation}. 

We use this two-site interaction to model the $S_1(\theta)$ type conformal interface, as they give the same $S$ matrix in the continuum limit. Therefore, the large $t$ behavior of its Loschmidt echo should match with our field theoretic prediction. 



\section{A Determinant Identity for the Boundary State Amplitude}
\label{app:pf_of_id}

In this appendix, we provide more details for calculating the amplitude $Z_{ab}$ in Sec.~\ref{sec:analytic_numerics}. We start to prove the following identity for a real symmetric matrix $M$
\begin{equation}
\label{eq:first_id_app_pf_of_id}
e^{- \vec{b}^{\dagger} M \vec{b} } e^{ \vec{b}^{\dagger} R \bar{\vec{b}}^{\dagger} }  = e^{ \vec{b}^{\dagger} e^{-M}  R \bar{\vec{b}}^{\dagger} } e^{- \vec{b}^{\dagger} M \vec{b} } ,
\end{equation}
where ${\bf b}$ and $\bar{\bf b}$ are vectors of bosonic operators. The matrix notation here should be understood as a bilinear expression as explained below Eq.~\eqref{eq:bd_state_matrix}.  

To prove Eq.~\eqref{eq:first_id_app_pf_of_id}, we first consider the special case where $R=\mathbb{I}$. We diagonalize $M = O^{T} \Lambda O $ and rotate the two sets of boson operators to the diagonal basis
\begin{equation}
  \vec{b}^{\dagger}  M \vec{b} = \vec{d}^{\dagger} \Lambda \vec{d}  \quad \vec{d} = O \vec{b} \quad \vec{\bar{d}}^\dagger = O^T \vec{\bar{b}}^\dagger,
\end{equation}
where we understand $\vec{\bar{b}}^\dagger$ as a column vector independent of $\vec{{b}}^\dagger$. Thus the whole expression can be written as
\begin{equation}
\begin{aligned}
  e^{- \vec{b}^{\dagger} M \vec{b} } e^{ \vec{b}^{\dagger} \bar{\vec{b}}^\dagger }  =  
  e^{- \vec{d}^{\dagger} \Lambda \vec{d} } e^{   \vec{d}^{\dagger} \bar{\vec{d}}^\dagger } = \prod_i  e^{- \lambda_i d_i^{\dagger} d_i } e^{  d_i^{\dagger} \bar{d}_i ^{\dagger} }.
\end{aligned}
\end{equation}
We recall for $ [X, Y] = sY $, 
\begin{equation}
  e^X e^{Y} = e^{\exp (s ) Y} e^{X},
\end{equation}
which is a solvable case of the Baker-Campbell-Hausdorff formula. Upon taking $X = -\lambda_i d_i^{\dagger} d_i$, $Y = d_i^{\dagger} \bar{d}^{\dagger}_i$, we have
\begin{equation}
\label{eq:lambda_commutator}
[- \lambda_i d_i^{\dagger} d_i, d_i ^{\dagger} \bar{d}_i^{\dagger}] =  - \lambda_i  d_i ^{\dagger} \bar{d}_i^{\dagger} ,
\end{equation}
and so $s = - \lambda_i$ for each $\lambda_i$. This enables us to commute those exponentials
\begin{equation}
\begin{aligned}
e^{- \vec{b}^{\dagger} M \vec{b} } e^{  \vec{b}^{\dagger} \bar{\vec{b}}^\dagger }   &= \prod_i e^{ e^{- \lambda_i }  d^{\dagger}_i \bar{d}^{\dagger}_i }  e^{-\lambda_i  d^{\dagger}_i d_i }  = e^{ \vec{b}^{\dagger} e^{-M}  \bar{\vec{b}}^\dagger} e^{- \vec{b}^{\dagger} M \vec{b} }. 
\end{aligned}
\end{equation}
For the general case where $R \neq\mathbb{I}$, we take $\vec{\bar{d}}^* = O^T R \vec{\bar{b}}^*$. This will not change the commutation relation of ${\bf d}$, and the role of $\bar{b}$ is decorative in Eq.~\eqref{eq:lambda_commutator}. Hence the rest of the proof follows the same way. \hfill$\blacksquare$

A direct consequence of Eq.~\eqref{eq:first_id_app_pf_of_id} is the following
\begin{equation}
\label{eq:second_id_in_app_pf_of_id}
Z_{ab} = \langle 0 | e^{ \vec{b} R_a^* \vec{\bar{b}} } e^{ - \vec{b}^{\dagger} M  \vec{b} } e^{  \vec{b}^{\dagger} R_b  \vec{\bar{b}}^{\dagger} }  |0  \rangle  = \frac{1}{\det( 1- R_a^{\dagger} e^{-M} R_b )} ,
\end{equation}
where $|0\rangle$ is the vacuum for ${\bf b}$ and $\bar{\bf b}$. 

One can use the identity in Eq.~\eqref{eq:first_id_app_pf_of_id} to reduce $Z_{ab}$ to 
\begin{equation}
Z_{ab} =   \langle 0 | \exp\Big\{ \vec{b} R_a^* \vec{\bar{b}}\Big\}  \exp \Big\{ \vec{b}^{\dagger} e^{-M}  R_b \bar{\vec{b}}^{\dagger}  \Big\}  |0 \rangle, 
\end{equation}
then a direct application of the MacMahon master theorem
\begin{equation}
  \langle 0 | \exp \Big\{ \vec{b}_1 X \vec{b}_2 \Big\}  \exp \Big\{ \vec{b}^{\dagger}_1 Y \vec{b}^{\dagger}_2 \Big\}|0  \rangle 
 = \frac{1}{\det(1 - X^T Y )}
\end{equation}
proves Eq.~\eqref{eq:second_id_in_app_pf_of_id}. \hfill$\blacksquare$



\section{Numerical Computation of Bipartite Fidelity and Loschmidt Echo}
\label{app:comp_fid_echo}

In this appendix, we provide technical details about the numerical calculation of the bipartite fidelity and Loschmidt echo.  Our strategy takes advantage of the symplectic structure of the bosonic Bogoliubov transformation and explicitly construct the "BCS" like ground state. With slight modification\cite{blaizot_quantum_1986}, one can work out its fermionic version and apply to quadratic fermion models in Ref.~\onlinecite{vasseur_universal_2014,stephan_local_2011}.

During the course of derivation in this and other appendices, we will repeatedly use the combinatorial identity called the McMahon master theorem
\begin{equation}
\label{eq:bosonic_McMahon}
\langle0|\exp\left\{\frac{1}{2}b_iX_{ij}b_j\right\}\exp\left\{\frac{1}{2}b^\dagger_iY_{ij}b^\dagger_j\right\}|0\rangle=\text{det}^{-\frac{1}{2}}(1-XY),
\end{equation}
for symmetric matrix $X$ and $Y$ and set of independent bosonic creation operators $b_i^{\dagger}$. One can prove it for (simultaneously) diagonalizable matrices and then claim its legitimacy for its combinatorial nature. 

\subsection{Boson Bogoliubov transformation}
\label{app_sub:boson_BdG}
We consider the following quadratic bosonic Hamiltonian
\begin{equation}
\begin{aligned}
\label{eq:quadratic_boson_H}
\hat{H} &= \frac{1}{2} (\vec{b}^{\dagger}, -\vec{b}) M 
\begin{pmatrix}
\vec{b}\\
\vec{b}^{\dagger} 
\end{pmatrix},  \qquad 
M = 
\begin{bmatrix}
A & -B^* \\
B & -A^* \\
\end{bmatrix},
\end{aligned}
\end{equation}
where ${\bf b}\equiv(b_{1},...,b_{n})^T$ is a vector of bosonic annihilation operators. The matrix $M$ consists of a $n\times n$ Hermitian block $A$ which plays the role of a single particle Hamiltonian in the fermionic case and symmetric block $B$ of the pairing interaction. 

We want to do a Bogoliubov transformation, which uses a $2n \times 2n$ matrix $S$ to define a diagonal basis 
\begin{equation}\begin{aligned}
\label{eq:def_a}
({\bf a} , {\bf a}^\dagger)\equiv({\bf b} , {\bf b}^\dagger)S
\end{aligned}\end{equation}
of the Hamiltonian. The transformation is canonical, meaning that it preserves the commutation relation 
\begin{equation}
\begin{aligned}
\label{eq:preserve_commutator}
J&\equiv\begin{bmatrix}
0 & \mathbb{I}\\
-\mathbb{I} & 0
\end{bmatrix}
=[
\begin{pmatrix}
{\bf a} \\
{\bf a}^\dagger
\end{pmatrix},
\begin{pmatrix}
{\bf a} & {\bf a}^\dagger
\end{pmatrix}]
=S^T[
\begin{pmatrix}
{\bf b} \\
{\bf b}^\dagger
\end{pmatrix},
\begin{pmatrix}
{\bf b} & {\bf b}^\dagger
\end{pmatrix}]S \\
&=S^T JS,
\end{aligned}
\end{equation}
where we have used the compact notation of the sort $([\vec{b}, \vec{b}^{\dagger}])_{ij} =  [b_i, b_j^\dagger]$ to denote the commutator matrix. The appearance of $J$ makes the symplectic nature of the problem manifest and we find $S$ is in the symplectic group ${\rm Sp}( 2n, \mathbb{C} ) $\cite{blaizot_quantum_1986,fulton_representation_2004}. Furthermore, the requirement that $a^\dagger$ is a complex conjugation of $a$ leads to the block structure of $S$
\begin{equation}
\label{eq:block_S}
S=
\begin{bmatrix}
u & v^*\\
v & u^*
\end{bmatrix}.
\end{equation}
And the blocks are constrained by the symplectic property
\begin{eqnarray}
  u^\dagger u-v^\dagger v&=\mathbb{I},\label{eq:constraint_1}\\
  u^T u-v^T v&=0\label{eq:constraint_2}.  
\end{eqnarray}
With these conditions, the Hamiltonian in basis $a$ becomes (the use of $({\bf b}^\dagger , -{\bf b})$ rather than $({\bf b}^\dagger , {\bf b})$ can be appreciated in this step)
\begin{equation}
H = \frac{1}{2} ( a^{\dagger}, -a )  (S^{\top} M (S^{\top})^{-1} )
\begin{pmatrix}
a\\
a^{\dagger} 
\end{pmatrix}.
\end{equation}
Quiet unusually, the diagonalization is performed by the symplectic group element.

To proceed, we introduce the real basis 
\begin{equation}\begin{aligned}
\label{eq:real_basis}
\begin{pmatrix}
{\bf b}\\
{\bf b}^\dagger
\end{pmatrix}
=C\begin{pmatrix}
\phi\\
\pi
\end{pmatrix}
=\frac{1}{\sqrt{2}}\begin{bmatrix}
1 & i \\
1 & -i
\end{bmatrix}\begin{pmatrix}
\phi\\
\pi
\end{pmatrix},
\end{aligned}\end{equation}
in which the Hamiltonian is
\begin{equation}\begin{aligned}
\hat{H}
&=\frac{1}{2}
\begin{pmatrix}
\phi & \pi
\end{pmatrix}
\begin{bmatrix}
\text{Re}(A-B^*) & -\text{Im}(A)+\text{Im}(B)\\
\text{Im}(A)+\text{Im}(B)& \text{Re}(A+B) \\
\end{bmatrix}
\begin{pmatrix}
\phi\\
\pi
\end{pmatrix} \\
&=\frac{1}{2}
\begin{pmatrix}
\phi & \pi
\end{pmatrix}
\mathcal{M}
\begin{pmatrix}
\phi\\
\pi
\end{pmatrix}.
\end{aligned}\end{equation}
It is not hard to check that $\mathcal{M}$ is real and symmetric. 

The general solution of the diagonalization problem is hard\cite{arnold_mathematical_2010}, however the positive definite $\mathcal{M}$ (and hence $M$) case can be solved by Williamson's theorem\cite{arnold_mathematical_2010,xiao_theory_2009,pirandola_correlation_2009,gosson_symplectic_2006}, which states the existence, uniqueness (up to reordering of eigenvalues) and explicit construction of the matrix $\mathcal{S}\in {\rm Sp}(2n \mathbb{R}) $ such that 
\begin{equation}\begin{aligned}
\mathcal{M}=\mathcal{S}
\begin{bmatrix}
\, d \, & \\
 & \, d\, \\
\end{bmatrix}
\mathcal{S}^T,
\end{aligned}\end{equation}
where the diagonal matrix $d$ are positive eigenvalues of $iJ\mathcal{M}$. After some algebra, we have
\begin{equation}\begin{aligned}
\label{eq:diagonalization_M}
M=J(C^{-1})^T\mathcal{S}C^TJ^{-1}
\begin{bmatrix}
d\\
&-d
\end{bmatrix}
C\mathcal{S}^TC^{-1}.
\end{aligned}\end{equation}
One can show that,
\begin{equation*}\begin{aligned}
S\equiv C\mathcal{S}^TC^{-1}
\end{aligned}\end{equation*}
is the required symplectic matrix in the complex basis. 

We will not elaborate on Williamson's theorem and its proof (see proofs in Ref.~\onlinecite{xiao_theory_2009,pirandola_correlation_2009,gosson_symplectic_2006} and also a recent application in the entanglement entropy context\cite{coser_contour_2017}). Instead we will show in App.~\ref{app_sub:harmonic_chain} that for the problem of the harmonic chain we are interested in, the diagonalization can be easily done without using the general recipe in the Williamson theorem.

\subsection{Groundstate in $b$ Basis}

Suppose we have obtained the required matrix $S$, the ground state will be the vacuum of the annihilation operators defined in Eq.~\eqref{eq:def_a} and in the $b$ basis it satisfies
\begin{equation}\begin{aligned}
\label{eq:a_vacuum_condition}
(b_iu_{ij}+b_i^\dagger v_{ij})|0\rangle_{{\bf a}}=0.
\end{aligned}\end{equation}
If the matrix $u$ is invertible, then we can introduce a matrix $T=vu^{-1}$ to rewrite Eq.~\eqref{eq:a_vacuum_condition} as
\begin{equation}\begin{aligned}
(b_i+b_j^\dagger T_{ji})|0\rangle_{{\bf a}}=0. 
\end{aligned}\end{equation}
The constraint Eq.~\eqref{eq:constraint_2} on the blocks of $u$ and $v$ (followed by the symplectic constraint of $S$) implies that $T$ is a symmetric matrix. With the observation of 
\begin{equation}\begin{aligned}
\exp\left\{-\frac{1}{2}b_j^\dagger T_{jk}b_k^\dagger\right\}b_i\exp\left\{\frac{1}{2}b_j^\dagger T_{jk}b_k^\dagger\right\}=b_i+T_{ij}b^\dagger_j,
\end{aligned}\end{equation}
we solve the groundstate
\begin{equation}\begin{aligned}
\label{eq:boson_BCS_gnd}
|0\rangle_{{\bf a}}&=\text{det}^{\frac{1}{4}}(1-T^\dagger T)\exp\left\{-\frac{1}{2}b_j^\dagger T_{jk}b_k^\dagger\right\}|0\rangle_{\bf b},\\
\end{aligned}\end{equation}
where the normalization is given by the McMahon master theorem Eq.~\eqref{eq:bosonic_McMahon}. Apply constraint in Eq.~\eqref{eq:constraint_1}, it simplifies to the top left corner of the symplectic matrix
\begin{equation}
\text{det}^{\frac{1}{4}}(1-T^\dagger T) =|\text{det}(u)|^{-\frac{1}{2}}.
\end{equation}

Eq.~\eqref{eq:boson_BCS_gnd} takes a similar form as the superconducting ground state, with the pairing wavefunction $T_{ij}$ determined by the Bogoliubov transformation. In the next section, we will see that the normalization factor gives the fidelity and Loschmidt echo.

\subsection{Boson fidelity} 
\label{app_sub:boson_fidelity}

Fidelity is defined as the (squared) overlap of groundstates of two different bosonic Hamiltonians. 

We start with a quadratic bosonic Hamiltonian $\hat{H}_0$ in the ${\bf b}$ basis, as in Eq.~\eqref{eq:quadratic_boson_H}. From the discussion in App.~\ref{app_sub:boson_BdG}, we are able to diagonalize it in the ${\bf a}$ basis for positive definite $M$. At $t=0$, the Hamiltonian becomes $\hat{H}_1$, which is still written in the ${\bf b}$ basis, but is diagonalized in a new basis ${\bf c}$. The corresponding Bogoliubov transformations read
\begin{equation}\begin{aligned}
\label{eq:two_BdG}
({\bf b} , {\bf b}^\dagger)S_0=({\bf a} , {\bf a}^\dagger),\quad
({\bf b} , {\bf b}^\dagger)S_1=({\bf c} , {\bf c}^\dagger),
\end{aligned}\end{equation}
and so
\begin{equation}\begin{aligned}
\label{eq:S0invS}
({\bf a} , {\bf a}^\dagger)=({\bf c} , {\bf c}^\dagger)\left(S_0^{-1}S_1\right)^{-1}.
\end{aligned}\end{equation}
One realizes that Eq.~\eqref{eq:S0invS} is another Bogoliubov transformation and so the corresponding matrix has the block structure
\begin{equation}
S_1^{-1}S_0=\begin{bmatrix}
u_1 & v_1^*\\
v_1 & u_1^*
\end{bmatrix}.
\end{equation}
Thus $|0\rangle_{\bf c}$ is related to the $|0\rangle_{{\bf a}}$ in the same way as in Eq.~\eqref{eq:boson_BCS_gnd}. Their overlap is therefore given by the normalization factor
\begin{equation}\begin{aligned}
|{}_{\bf a}\langle0|0\rangle_{\bf c}|^2=\frac{\Big|{}_{\bf a}\langle 0 | \exp( -\frac{1}{2} a_j^{\dagger} T^{jk} a_k^{\dagger} )|0   \rangle_{\bf a} \Big|^2}{|\text{det}(u_1)|} = \frac{1}{|\text{det}(u_1)|}.
\end{aligned}\end{equation}

\subsection{Boson Loschmidt echo}
\label{app_sub:boson_Loschmidt_echo}

The Loschmidt echo is defined as the (squared) overlap of the evolved state
\begin{equation}\begin{aligned}
|0\rangle_{{\bf a}(t)}\equiv e^{-i\hat{H}_1t}|0\rangle_{\bf a},
\end{aligned}\end{equation}
with $|0\rangle_{\bf a}$ the ground state of the Hamiltonian $\hat{H}_0$ before the quench. We introduce a dynamical basis
\begin{equation}\begin{aligned}
a_i(t)&=e^{-i\hat{H}_1t}a_ie^{i\hat{H}_1t},
\end{aligned}\end{equation}
which annihilate the evolved state at time $t$: $a_i(t)|0\rangle_{{\bf a}(t)}=0$. Upon using the diagonal basis $\hat{H}_1=\sum_iE_ic_i^\dagger c_i$, the Bogoliubov transformation at time $t$ can be represented as a chain of symplectic transformation
\begin{equation}\begin{aligned}
\label{eq:BdG_transformation_echo}
&({\bf a}(t),{\bf a}^\dagger(t))=e^{-iHt}
({\bf a} , {\bf a}^\dagger)e^{iHt} \\
=&({\bf a},{\bf a}^\dagger)S_0^{-1}S_1\text{diag}(e^{iEt},e^{-iEt})S_1^{-1}S_0.
\end{aligned}\end{equation}
It is evident that the evolved state $|0\rangle_{{\bf a}(t)}$ is related to the $|0\rangle_{{\bf a}}$ in the same way as in Eq.~\eqref{eq:boson_BCS_gnd}. The overlap, as we have seen in the fidelity case, is the normalization factor of the "BCS" ground state. It is related to the top left block of the Bogoliubov transformation in Eq.~\eqref{eq:BdG_transformation_echo},
\begin{equation}\begin{aligned}
\label{eq:echo_t}
\mathcal{L}(t)=|_{\bf a}\langle0|0\rangle_{{\bf a}(t)}|^2=|\text{det}(u_1^\dagger e^{iEt}u_1-v_1^\dagger e^{-iEt}v_1)|^{-1}.
\end{aligned}\end{equation}

\subsection{Harmonic chain} 
\label{app_sub:harmonic_chain}

In this subsection, we explicitly construct the matrix $\mathcal{S}$ for the case of the harmonic chain introduced in Sec.~\ref{sec_sub:free_boson_lattice}. In the basis defined in Eq.~\eqref{eq:real_basis}, the Hamiltonian for 1D harmonic chain is
\begin{equation}\begin{aligned}
\hat{H}
=\frac{1}{2}
\begin{pmatrix}
\phi & \pi
\end{pmatrix}
\mathcal{M}
\begin{pmatrix}
\phi\\
\pi
\end{pmatrix}
=\frac{1}{2}
\begin{pmatrix}
\phi & \pi
\end{pmatrix}
\begin{bmatrix}
\mathcal{V} \\
& {\mathbb{ I}}
\end{bmatrix}
\begin{pmatrix}
\phi\\
\pi
\end{pmatrix},
\end{aligned}\end{equation}
where $\mathcal{V}$ is real symmetric matrix that can be diagonalized as $\mathcal{V}=\mathcal{O}D^2\mathcal{O}^T$. The matrix $\mathcal{V}$ depends on the boundary condition, but positive definiteness is the only requirement here. 

The matrix $\mathcal{S}$ that diagonalizes $\mathcal{M}$
\begin{equation}
\begin{aligned}
\mathcal{M}=\mathcal{S}
\begin{bmatrix}
D \\ 
& D
\end{bmatrix}
\mathcal{S}^T
\end{aligned}
\end{equation}
is given by the following real symplectic matrix
\begin{equation}\begin{aligned}
\mathcal{S}\equiv
\begin{bmatrix}
\mathcal{O}D^{1/2} \\
& \mathcal{O}D^{-1/2}
\end{bmatrix}.
\end{aligned}\end{equation}

Thus the Hamiltonian is diagonalized as
\begin{equation}\begin{aligned}
M=S^{-1}
\begin{bmatrix}
D \\
& -D
\end{bmatrix}
S,
\end{aligned}\end{equation}
where the Bogoliubov transformation take the desired block form
\begin{equation}\begin{aligned}
S&=C\mathcal{S}C^{-1}
=\begin{bmatrix}
O(D^{1/2}+D^{-1/2}) & O(D^{1/2}-D^{-1/2}) \\
O(D^{1/2}-D^{-1/2}) & O(D^{1/2}+D^{-1/2}) 
\end{bmatrix}.
\end{aligned}\end{equation}



\bibliographystyle{unsrt}
\bibliography{bCFT_PRB}

\end{document}